\documentclass[paper]{JHEP3}
\usepackage{graphicx}
\newcommand{\be}{\begin{equation}}
\newcommand{\ee}{\end{equation}}
\newcommand{\ba}{\begin{eqnarray}}
\newcommand{\ea}{\end{eqnarray}}

\preprint{
LU TP 07-26\\
arXiv:0709.0230 [hep-ph]\\
Revised October 2007}

\title{$\eta \to 3 \pi$ at Two Loops In Chiral Perturbation Theory}

\author{
Johan Bijnens
and 
Karim Ghorbani
\\
Department of Theoretical Physics, Lund University,\\
S\"olvegatan 14A, S 223-62 Lund, Sweden\\
Email: \email{bijnens@thep.lu.se}, \email{karim.ghorbani@thep.lu.se} }

\abstract{
We calculate the decay $\eta\to3\pi$ at next-to-next-to-leading order or
order $p^6$ in Chiral Perturbation Theory. The corrections are somewhat larger
than was indicated by dispersive estimates. 
We present numerical results for the Dalitz plot parameters, the
ratio $r$ of the neutral to charged decay and the total decay rate.
In addition we derive an inequality between the slope parameters
of the charged and neutral decay.
The experimental charged decay rate leads
to central values for the isospin breaking quantities
$R=42.2$ and $Q=23.2$.
}


\begin{document}

\section{Introduction}
\label{sect:introduction}

Since its discovery, the eta particle has been under 
tense scrutiny, fairly recent reviews are \cite{etahandbook,eta05}.
The eta decay to three pions is particularly interesting.
It can only happen due to isospin breaking. 
This implies 
that the decay rate vanishes in the 
limit of equal of up and down quark-masses, 
ignoring electromagnetic effects.
The very first attempts to explain the decay 
\cite{Sutherland1,Sutherland2},
considering it to be an electromagnetic decay,
resulted in an almost zero decay rate, which was 
in flat disagreement with experiment. 

Later on, a combination of current algebra 
techniques in SU(3) and the partially conserved 
axial-vector current (PCAC) hypothesis could make a better 
prediction~\cite{orderp2x1,orderp2x2}.
This, however, underestimated the observed decay rate by a factor of a few.
PCAC and current algebra were generalized into Chiral Perturbation Theory
(ChPT)
and brought into a modern form \cite{Weinberg0,GL0,GL1}.
The lowest order for $\eta\to 3\pi$ was calculated in
\cite{orderp2x1,orderp2x2} and the next-to-leading order (NLO) in \cite{GL3}.
The main goal of this paper is to 
obtain next-to-next-to-leading-order (NNLO) expression for
this decay.

We give now a short overview of the present situation, closely following
the discussion in \cite{BG}.

The well-known tree-level result for this decay 
channel can be derived from current algebra or ChPT
and has the structure
\be
A(s,t,u) =\frac{ B_0 (m_{u} - m_{d}) }{3\sqrt{3} F_{\pi}^2 } 
\left( 1+\frac{3(s-s_{0})}{m^{2}_{\eta}-m^{2}_{\pi} } \right)\,.  
\ee
The prefactor $m_{u} - m_{d}$ shows that the decay is isospin violating
or $SU(2)_V$-violating. The magnitude of $m_u-m_d$
determines the size of the isospin
symmetry breaking coming from the strong interaction itself.
A precise determination of this quantity is in principle possible here
because its size has a direct impact on the decay rate. 

Using Dashen's theorem this factor 
can be obtained from the physical meson masses
by removing electromagnetic effects to lowest order.
Following the line of the argument outlined in 
Dashen's theorem \cite{Dashen} 
one arrives at 
\be
 B(m_{d} - m_{u}) = m^{2}_{K^{0}} - m^{2}_{K^{+}}- 
m^{2}_{\pi^{0}} + m^{2}_{\pi^{+}}\,.  
\ee
Under these circumstances, the theoretical decay 
width is 66 eV to be compared with 295~eV 
from experiment~\cite{PDG}.
One may consider three potential 
sources for this discrepancy. First, the violation 
of the Dashen's theorem due to the electromagnetic 
interferences increases the value of 
$m_{d}-m_{u}$ \cite{Dashen1,Dashen2} and therefore 
the decay width, but this is not sufficient
as will be discussed more in the discussion section.
Secondly, electromagnetic corrections of the decay 
amplitude which are of higher order, order $e^2p^2$, are safely
negligible in comparison with the strong interactions
as pointed out in \cite{Kambor0} where the analysis of
\cite{Sutherland1,Sutherland2} was brought to NLO.
   
Finally the contribution of the higher order chiral effects 
must be taken into account. Especially since the strong $\pi \pi$ 
rescattering in the $S$-wave channel may develop a significant 
correction, see e.g. \cite{unitarity1,unitarity2}. The NLO corrections were
obtained in \cite{GL3}.
The unitarity 
correction at one-loop level is about half of the 
total NLO effects. This, of course, confirmed
the fact that by virtue of a large eta mass,
significant rescattering effects in the $S$-wave
can occur.
The vertex 
corrections and tree graphs make up the rest 
of the NLO contribution. 
The coupling 
constants involved at this level can be rewritten in terms of meson masses
except for $L_{3}^r$. In \cite{GL0,GL3} $L_3^r$ was estimated
by invoking the OZI (Okubo-Zweig-Iizuka) rule and comparing with $\pi\pi$
scattering lengths.
Their finding for the 
decay rate of $\eta\to3\pi$ is $160\pm{50}$ which is still 
far from the experimental value, however with 
a large theoretical error.

Given the importance of the unitarity correction at NLO, it was
deemed necessary to estimate this part of the corrections
to higher order. This can be done using dispersive methods.
In \cite{Kambor1}, 
extended Khuri-Treiman equations are used to 
evaluate the two-pion rescattering to the decay 
$\eta \to \pi \pi \pi $. They achieve a moderate 
modification, an increase of about 14\% per cent in the amplitude at
the center of the Dalitz plot.
Moreover, another analysis,  based on a somewhat different
dispersive method, but also restricting itself to two-pion rescattering,
represented in \cite{Anisovich1} suggests 
also a mild enhancement to the real part of the 
amplitude in the physical region. A more model-dependent
analysis of dispersive corrections
appeared recently \cite{Borasoy} relying on combining
$U(3)\times U(3)$ ChPT 
and a relativistic coupled-channels method, finds
agreement with data.

Given all these, our motivation to perform a full 
NNLO computation is twofold.
First, due to a relatively large strange 
quark-mass, the convergence of three-flavour or $SU(3)$ ChPT
is an a priori question. The reason hinges on the fact 
that the ratio  
${M_{K}^{2}}/{M_{\rho}^{2}}$ is much larger than
${M_{\pi}^{2}}/{M_{\rho}^{2}}$ and there are
possibly large effects as a result of 
strange quark loops. Three-flavour ChPT is probably less convergent
than two-flavour ChPT, see e.g. \cite{Paris} for a discussion. 
In general, one needs to have several terms available in order
to check convergence. The situation at present is not fully clear,
The results for the vector form factors, $K_{\ell4}$ and $\pi\pi$-scattering
have an acceptable convergence,
while the results for the masses and $\pi K$-scattering,
seem to converge slower, see~\cite{reviewp6} and references therein.
Therefore we would like to check explicitly 
whether one may treat the strange 
quark-mass perturbatively for this
process, namely $\eta \to 3\pi$.
In addition, at NLO the unitarity correction provided only half of the
total correction. It is therefore also of interest to know if the
other corrections are important at NNLO as well. This is known to be the
case for $K_{\ell4}$ by comparing \cite{BCG} and \cite{ABT3}.
 
Our finding shows that the full amplitude up to 
and including order $p^{6}$
corrections converges reasonably acceptably but we find larger corrections
than in \cite{Kambor1,Anisovich1}.

In this paper, we perform the full NNLO calculation of $\eta\to3\pi$
in standard three-flavour ChPT. We do this to first order in the
isospin breaking quantity $m_u-m_d$. In addition,
we perform a first numerical analysis with this expression.
We have therefore also estimated the order $p^6$ coupling constants $C_{i}^r$, 
using a resonance chiral Lagrangian, assuming that
vector and scalars mesons saturate the $C_{i}^r$. In addition, we derive an
inequality between the slope parameters.

The layout of this article is as follows.
In Sect.~\ref{sect:chpt} a brief introduction to
Chiral Perturbation Theory is provided. 
Sect.~\ref{sect:amplitude} describes how to calculate the $\eta\to3\pi$ 
amplitude in the presence of mixing, the kinematics for the decay
and the form of the amplitude at NNLO. We also show the Feynman diagrams
and provides references to
how we deal with the loop integrals and renormalization. A short discussion
on our analytic results is given in Sect.~\ref{sect:analytical}
followed by our estimate of order $p^6$ low energy constants in
Sect.~\ref{sect:reso}.
A main part of the manuscript is the numerical results and comparison with
experiment given Sect.~\ref{sect:numerical}. We first present a discussion
on the amplitude level, Sect.~\ref{sect:firstlook}, then compare with the
earlier dispersive results, Sect.~\ref{sect:dispersive}.
The main comparison with experiment is done in Sect.~\ref{sect:dalitz} for
the Dalitz plot distributions and in Sect.~\ref{sect:decay} for the ratio
of amplitudes. We present the results for the value of $R$ and $Q$
in Sect.~\ref{sect:RQ}. The App.~\ref{app:dalitz} contains a discussion
on the cancellations inherent in $\alpha$ and the derivation of the
inequality between the slope parameters. The remaining appendices
present our NLO expression and the
dependence on the order $p^6$ low energy constants.
 
\section{Chiral Perturbation Theory}
\label{sect:chpt}

One approach to address Quantum Chromodynamics (QCD) 
at low energy is the application of effective field theories.
In \cite{Manohar} and references therein some basic 
concepts and a few interesting examples can be found for this method.
As an effective field theory, ChPT
is constructed based on the approximate chiral symmetry of the underlying 
theory (QCD) and is an expansion in external momenta and quark-masses,
momenta and meson masses are generically denoted by $p$ and the
expansion is in powers of these.
The dynamical degrees of freedom i.e. pseudo-Goldstone particles, 
are manifested as a result of the spontaneous chiral symmetry
breaking of QCD. Weinberg \cite{Weinberg0}
systematized the use of effective field theory as 
an alternative to the current algebra formalism,
incorporating naturally the chiral logarithms. 
Gasser and Leutwyler in two elegant papers presented this expansion including 
terms of order $p^{4}$ and introduced the external field method \cite{GL0}.
They also formulated the
extension to three light flavours \cite{GL1}.
They found a substantial correction to the $\pi \pi$ scattering 
lengths and effective ranges at this order.
Reviews of ChPT at order $p^4$ are \cite{Bernard1,Eckerreview}.
This line of work has been further developed to include $p^{6}$ 
corrections, see the review \cite{reviewp6}.
A more introductory recent review is \cite{Bernard2}.

The full action consists of subterms with a definite number of derivatives 
or powers of quark-masses as shown below
\be
\label{lagL2}
\mathcal{L}_{eff} = \mathcal{L}_{2} + \mathcal{L}_{4} + \mathcal{L}_{6}
 + \cdots
\,.
\ee
The subscript indicates the chiral order. Quark masses are 
counted as order $p^{2}$ using the lowest order relation
$m_\pi^2 = B_0\left(m_u+m_d\right)$. The lowest order chiral Lagrangian 
incorporates two parameters and has the form
\be
\mathcal{L}_{2} = \frac{F_0^{2}}{4} \left( \langle D_{\mu}U D^{\mu}U^{\dagger}
 \rangle + 
\langle\chi U^{\dagger} + U \chi^{\dagger}  \rangle  \right) 
\ee
and the next-to-leading Lagrangian or order $p^4$ Lagrangian
is given as\cite{GL1}
\ba
\label{lagL4}
{\cal L}_4&&\hspace{-0.5cm} = 
L_1 \langle D_\mu U^\dagger D^\mu U \rangle^2
+L_2 \langle D_\mu U^\dagger D_\nu U \rangle 
     \langle D^\mu U^\dagger D^\nu U \rangle \nonumber\\&&\hspace{-0.5cm}
+L_3 \langle D^\mu U^\dagger D_\mu U D^\nu U^\dagger D_\nu U\rangle
+L_4 \langle D^\mu U^\dagger D_\mu U \rangle
 \langle \chi^\dagger U+\chi U^\dagger \rangle
\nonumber\\&&\hspace{-0.5cm}
+L_5 \langle D^\mu U^\dagger D_\mu U (\chi^\dagger U+U^\dagger \chi ) \rangle
+L_6 \langle \chi^\dagger U+\chi U^\dagger \rangle^2
\nonumber\\&&\hspace{-0.5cm}
+L_7 \langle \chi^\dagger U-\chi U^\dagger \rangle^2
+L_8 \langle \chi^\dagger U \chi^\dagger U
 + \chi U^\dagger \chi U^\dagger \rangle
\nonumber\\&&\hspace{-0.5cm}
-i L_9 \langle F^R_{\mu\nu} D^\mu U D^\nu U^\dagger +
               F^L_{\mu\nu} D^\mu U^\dagger D^\nu U \rangle
\nonumber\\&&\hspace{-0.5cm}
+L_{10} \langle U^\dagger  F^R_{\mu\nu} U F^{L\mu\nu} \rangle
+H_1 \langle F^R_{\mu\nu} F^{R\mu\nu} + F^L_{\mu\nu} F^{L\mu\nu} \rangle
+H_2 \langle \chi^\dagger \chi \rangle\,.
\ea
The matrix $U \in SU(3)$ parameterizes 
the octet of light pseudo-scalar mesons with its exponential
representation given in terms of mesonic fields 
matrix as 
\be
U(\phi) = \exp(i \sqrt{2} \phi/F_0)\,,
\ee
where
\ba
\phi (x) 
 = \, \left( \begin{array}{ccc}
\displaystyle\frac{ \pi_3}{ \sqrt 2} \, + \, \frac{ \eta_8}{ \sqrt 6}
 & \pi^+ & K^+ \\
\pi^- &\displaystyle - \frac{\pi_3}{\sqrt 2} \, + \, \frac{ \eta_8}
{\sqrt 6}    & K^0 \\
K^- & \bar K^0 &\displaystyle - \frac{ 2 \, \eta_8}{\sqrt 6}
\end{array}  \right) .
\ea
The covariant derivative and the field strength tensor 
are defined as
\be
D_\mu U = \partial_\mu U -i r_\mu U + i U l_\mu \,, \quad
F_{\mu\nu}^{L} = \partial_\mu l_\nu -\partial_\nu l_\mu
-i \left[ l_\mu , l_\nu \right]\,,
\ee
and a similar definition for the right-handed field strength.
Here $l_\mu$ and $r_\mu$ represents the left-handed and 
right-handed chiral currents respectively. 
$\chi$ is parameterized in terms of scalar ($s$) and pseudo scalar ($p$) 
external densities
as $\chi = 2B_0 \left( s + i p\right)$. In the process discussed in this article
we have set $s = diag ( m_{u}, m_{d}, m_{s} )$ and $p = l_\mu = l_\nu = 0$.
Finally, the notation $\langle A\rangle = \mathrm{Tr}_F\left(A\right)$,
the trace over flavours.
The $SU(3)$ chiral Lagrangian of order $p^6$ contains 94 operators. For its 
explicit form we refer to \cite{BCE1}.  


Tree diagrams using only vertices from the Lagrangian $\mathcal{L}_{2}$, 
provide the lowest order term in the expansion.
In general, tree-level diagrams
make up the semiclassical part of the unitarity of the S-matrix.
We thus ought to include loop effects. 
The infinities which arise from one loop 
diagrams with vertices taken from $\mathcal{L}_{2}$ 
can not be absorbed by renormalizing $F_0$ and $B_0$ 
or rescaling the fields since these contribute at tree level
at a \emph{different} order in $p^2$ from the one-loop diagrams.
Renormalization at higher orders contains many subtleties. The procedure
used here is discussed extensively in \cite{BCEGS2,BCE2}. 
The full divergence structure at order $p^4$ \cite{GL0,GL1}
and $p^6$ \cite{BCE2} is known.
For our calculation,
the cancellation of the divergences
is an important cross-check.      
The nontrivial predictions of ChPT, the so called chiral logarithms,
are due to infrared singularities
in the chiral limit due to the (pseudo)-Goldstone boson intermediate states.
Lattice QCD computations support this 
logarithmic behavior when compared with ChPT results 
evaluated in finite volume \cite{Aubin:2004fs}. 

The part not determined by ChPT, the $L_i^r$ and higher order LECs, $C_i^r$
 at order $p^6$, encode the information about higher
scale physics which has no dynamical role in our 
effective field theory.
A detailed discussion on the determination of the order $p^4$ LECs is given 
in \cite{GL1} at order $p^4$ and at order $p^6$ in \cite{reviewp6,ABT3,ABT4}.

A systematic extension to effective field 
theories incorporating the resonance fields may provide a 
profound theoretical ground to ultimately underpin  
the values of the LECs. Resonance field methodology 
takes its original form in Sakurai's hypothesis of vector 
meson dominance. It was worked out at order $p^4$ by \cite{EGPR}.
We will only use it for order $p^6$ LECs in the simplified form
discussed in Sect.~\ref{sect:reso}.
More systematic approaches at order $p^6$ exist \cite{Cirigliano1}
but there are also caveats to be observed from short-distance
constraints, both positive and negative~\cite{EGLPR,BGLP}.

\section{Eta decay amplitude: formalism}
\label{sect:amplitude}

\subsection{Matrix-elements in the presence of mixing}
\label{sect:mixing}

In this section we explain how to calculate matrix-elements
by the use of the Lehmann-Symanzik-Zimmermann (LSZ) 
reduction formula in the present of mixing.
This is a generalization of the discussion in \cite{ABT4} Sect. 2,
to the case needed here.
\FIGURE{
\includegraphics[width=0.3\textwidth,clip]{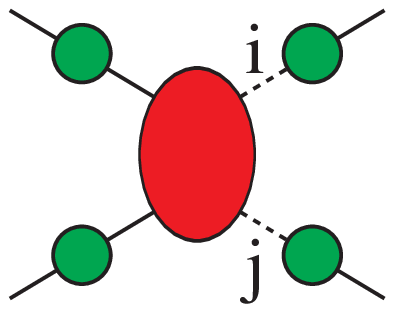}
\caption{
The full four-point Green function is represented. 
The Oval stands for the amputated four-point Green function and
circles indicate the full two-point functions. 
The solid lines are external mesons 
and the dashed lines labeled by i and j, indicate the sum
over states implied in the two external legs where mixing occurs.}
\label{fig:green_function}
}
The scattering amplitude is basically the residue of
the Fourier transformed Green function in the limit where
all the outgoing particles go on-shell.

For the case of $\eta \to \pi^+ \pi^- \pi^0$, 
the mixing occurs in the two external legs involving neutral particles
pion and eta as illustrated in Fig.~\ref{fig:green_function}.
For the decay $\eta\to\pi^0\pi^0\pi^0$ mixing is relevant in all four
external legs. In \cite{ABT4} two-point functions were analyzed in all
generality as well as amplitudes where one external leg could undergo mixing.
Here, we reiterate some basic ingredients introduced 
there and will lead to derive a general formula that relates the 
amputated four-point Green function to 
the scattering amplitude order by order in 
perturbative expansion of $\eta\to\pi^+\pi^-\pi^0$.
In the ChPT calculation we only retain terms to first order in isospin
breaking, we therefore can use the relation (\ref{isorel}) and do not need
a more general formula for $\eta\to\pi^0\pi^0\pi^0$ to all orders in the mixing.

The amplitude or matrix-element for 
a process with $n$ ingoing and outgoing 
particles can be expressed in the form   
\be
\label{defLSZ}
{\cal A}_{i_1\ldots i_n}
= 
\left( \frac{(-i)^{n}}{\sqrt{Z_{i_1} \ldots Z_{i_n}}} \right)\, 
\prod_{i=1}^n \, 
\lim_{k_i^2\to m_i^2}(k_i^2-m_i^2)\,
G_{i_1\ldots i_{n}}
(k_1,\ldots,k_n)\,.
\ee
The function $G_{i_1\ldots i_{n}}(k_1,\ldots,k_n)\, $ is the 
exact $n$-point Green function\footnote{We consider as operators that
are in the Green functions the fields as present in the Lagrangian.
The formula is valid for all operators with a nonzero coupling to
single-particle states.}
and the coefficients $Z_{i}$
are defined as
\be
G_{ii}(p^2\approx m_{i\,phys}^2) = \frac{i Z_i}{p^2-m_{i\,phys}^2}\,.
\ee
These are often called field-strength or wave-function renormalization factors.
 
We begin by decomposing the full four-point function into 
the amputated four-point function and four full propagators or two-point
functions for the external legs.  
\ba
\label{decomposition}
G_{1238}  &=&  G_{11}(p^2\approx m_{1\,phys}^2) 
G_{22}(p^2\approx m_{2\,phys}^2) 
\nonumber\\&&
G_{3i}(p^2\approx m_{3\,phys}^2)G_{8j}(p^2\approx m_{8\,phys}^2) 
{\cal G}_{12ij}\,.
\ea
Subscripts in $G_{1238}$ are designated for the four external
particles\footnote{We use a somewhat generic notation here, 1,2 for the
nonmixing external legs and 3,8 for the mixing external legs corresponding to
$\pi^0$ and $\eta$.}
namely, $\pi^1 ,\pi^2, \pi^3 $and $\pi^8$. A summation over the possible values
of $i,j$ is implied in (\ref{decomposition}).
Notice that what we observe 
as particles are those in the physical basis. The numbering
in (\ref{decomposition}) and the rest corresponds to the labeling
of the fields in the Lagrangian.
We express the amputated Green function ${\cal G}_{12ij}$
in terms of the contributions at the different chiral orders:
\be
{\cal G}_{12ij}  = {\cal G}^{(2)}_{12ij}+ 
{\cal G}^{(4)}_{12ij}+ {\cal G}^{(6)}_{12ij}+\cdots\,.  
\ee

In view of the fact that only the two fields associated with the 
neutral pion and eta particles mix, the relevant 
two-point functions $G_{ij}$ constitute a two-by-two matrix
$G=\left(G_{ij}\right)$
with $i,j=3,8$. This can be done since
the only nonzero off-diagonal elements are $G_{38}=G_{83}$.
For this matrix
we define $G^{-1} = -i{\cal P}$ and ${\cal P}$ can be written as
${\cal P} = P^{-1}+\Pi$.
$P^{-1} = \mathrm{diag}(p^{2}-m_{i0}^{2})$ is the matrix of (inverse)
lowest order propagators. We \emph{assume} here that the fields in the
Lagrangian have been diagonalized to lowest order.
The quantity $\Pi$ denotes the sum of all the
one-particle-irreducible diagrams as shown pictorially below.
\centerline{\includegraphics{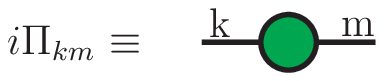}}
The function G in matrix form is as follows 
\be
G = G(p^2) = \frac{i}{{\cal P}_{33}(p^2)\,{\cal P}_{88}(p^2)
 - {\cal P}_{38}^2(p^2)}
\left( \begin{array}{cc}
{\cal P}_{88}(p^2)   & -{\cal P}_{38}(p^2)  \\
-{\cal P}_{38}(p^2)  & {\cal P}_{33}(p^2) 
    \end{array} \right)\,
\ee
The denominator is in fact $\det{\cal P}(p^2)$
and vanishes at the physical masses given by poles in $G$.
One may exploit $det{\cal P}(p^2)=0$ to 
obtain the perturbative corrections to the bare mass \cite{ABT4}.    
By expanding the denominator around the physical mass 
and knowing that $\det$ ${\cal P}(p^2)$
vanishes at the limit of physical mass, we can then identify 
the field-strength factors for the neutral pion and eta as 
\be
Z_{3} = \frac{1}
{\left.\frac{\partial}{\partial p^2}
\left(\det{\cal P}(p^2)\right)\right|_{p^2=m_\pi^2}} {\cal P}_{88}(m_\pi^2)\,,
\quad
Z_{8} = \frac{1}
{\left.\frac{\partial}{\partial p^2}
\left(\det{\cal P}(p^2)\right)\right|_{p^2=m_\eta^2}} 
{\cal P}_{33}(m_\eta^2)\,.
\ee
We define
\be
Z_{ii} = \left.\frac{\partial\Pi_{ii}}{\partial p^2}
\right|_{m{}^2_{i\mathrm{phys}}}
\ee
and expand also the remaining quantities by chiral order
\be
\Pi_{ij} = \Pi^{(4)}_{ij} +\Pi^{(6)}_{ij}+\cdots\,,
\quad
Z_{ii} = 1+ Z^{(4)}_{ii} + Z^{(6)}_{ii}+\cdots\,.
\ee
Note that by definition $\Pi$ only starts at NLO in the chiral expansion.

We now have all the ingredients to put in (\ref{defLSZ}) and perform the
chiral expansion for the amplitude. We obtain
using $Z_{11}=Z_{22}$ and $\Pi_{38}(p^2)=\Pi_{83}(p^2)$
up to order $p^6$:
\ba
\label{fullamplitude}
{\cal A}_{1238} &=& {\cal A}_{1238}^{(2)}+
      {\cal A}_{1238}^{(4)}+{\cal A}_{1238}^{(6)}+\cdots\,,
\\
\label{fullp2}
{\cal A}_{1238}^{(2)} &=& {\cal G}_{1238}^{(2)} \,,
\\
\label{fullp4}
{\cal A}_{1238}^{(4)} &=&
{\cal G}_{1238}^{(4)} 
- \left( Z_{11}^{(4)} + \frac{1}{2} Z_{33}^{(4)} + 
 \frac{1}{2} Z_{88}^{(4)} \right){\cal G}_{1238}^{(2)}
-\frac{\Pi_{38}^{(4)}}{\Delta m_1^2} {\cal G}_{1288}^{(2)} 
-\frac{\Pi_{38}^{(4)}}{\Delta m_2^2} {\cal G}_{1233}^{(2)} \,,
\\
\label{fullp6}
{\cal A}_{1238}^{(6)} &=&
{\cal G}_{1238}^{(6)}
-\frac{1}{2} \left(2 Z_{11}^{(6)}+ Z_{33}^{(6)}+ Z_{88}^{(6)} \right)
        {\cal G}_{1238}^{(2)}
-\frac{1}{2} \left(2 Z_{11}^{(4)}+ Z_{33}^{(4)}+ Z_{88}^{(4)} \right)
      {\cal G}_{1238}^{(4)}
\nonumber\\&&
+\frac{3}{8} \left( \left( Z_{33}^{(4)}\right)^2 +\left( Z_{88}^{(4)}\right)^2 
      \right) {\cal G}_{1238}^{(2)} 
+\left( Z_{33}^{(4)}\right)^2 {\cal G}_{1238}^{(2)}
\nonumber\\&& 
+\left( \frac{1}{4}Z_{33}^{(4)}\, Z_{88}^{(4)}
        + \frac{1}{2}Z_{11}^{(4)}\, Z_{11}^{(4)}
        + \frac{1}{2}Z_{11}^{(4)}\, Z_{11}^{(4)} \right)
     {\cal G}_{1238}^{(2)}
\nonumber\\&&
-\frac{\Pi_{38}(3)^{(4)}}{\Delta m_1^2}{\cal G}_{1288}^{(4)}
-\frac{\Pi_{38}(8)^{(4)}}{\Delta m_2^2}{\cal G}_{1233}^{(4)}
-\frac{\Pi_{38}(3)^{(6)}}{\Delta m_1^2}{\cal G}_{1288}^{(2)}
-\frac{\Pi_{38}(8)^{(6)}}{\Delta m_2^2}{\cal G}_{1233}^{(2)}
\nonumber\\&&
+\frac{\Pi_{38}(3)^{(4)}\,\Pi_{88}(3)^{(4)}}{\Delta m_1^2}{\cal G}_{1288}^{(2)}
+\frac{\Pi_{38}(8)^{(4)}\,\Pi_{88}(3)^{(4)}}{\Delta m_2^2}{\cal G}_{1233}^{(2)}
\nonumber\\&&
+\frac{1}{2} \left( Z_{33}^{(4)}\frac{\Pi_{38}(3)^{(4)}}{\Delta m_1^2}
                   +Z_{88}^{(4)}\frac{\Pi_{38}(3)^{(4)}}{\Delta m_1^2} 
                   +Z_{11}^{(4)}\frac{\Pi_{38}(3)^{(4)}}{\Delta m_1^2}
            \right)  {\cal G}_{1288}^{(2)}
\nonumber\\&&
+\frac{1}{2} \left(  Z_{33}^{(4)}\frac{\Pi_{38}(8)^{(4)}}{\Delta m_2^2}
                    +Z_{88}^{(4)}\frac{\Pi_{38}(8)^{(4)}}{\Delta m_2^2}
                    +Z_{11}^{(4)}\frac{\Pi_{38}(8)^{(4)}}{\Delta m_2^2}
              \right)  {\cal G}_{1233}^{(2)}
\,.
\ea
We defined the abbreviations $\Delta m^{2}_{1} = m^2_{\pi}- m^2_{\eta_{0}}$
and $\Delta m^{2}_{2} = m^2_{\eta}- m^2_{\pi_{0}}$ with first a physical mass
and the second term the lowest order mass.
$\Pi_{ij}^{(k)}(I)$ are evaluated at the physical $\pi^0$ mass for $I=3$ and at
the physical $\eta$ mass for $I=8$.
The ${\cal G}_{12ij}^{(n)}$
are evaluated at the physical charged pion mass for the legs with indices
1 or 2 and at the physical $\pi^0$ mass for the leg labeled $i$ and
at the physical $\eta$ mass at the leg labeled $j$.
The set of terms 
are shown at the different orders.

\subsection{Kinematics and isospin}
\label{sect:defamplitude}

We write the amplitude for the decay
 $\eta(p_\eta)\to\pi^+(p_+)\pi^-(p_-)\pi^0(p_0)$
using the Mandelstam variables
\ba
s &=& (p_+ +p_-)^2  = (p_\eta - p_0)^2   \,,
\nonumber\\
t &=& (p_+ +p_0)^2   = (p_\eta -p_-)^2     \,,
\nonumber\\
u &=& (p_- +p_0)^2   = (p_\eta -p_+)^2      ,.
\ea
These are linearly dependent
\be
s+t+u = m_{\pi^{o}}^2 + m_{\pi^{-}}^2 + m_{\pi^{+}}^2 + m_{\eta}^2
\equiv 3 s_0\,.  
\ee
Due to the $SU(2)_V$ symmetry breaking the isospin
basis and physical basis in the $\pi^0$-$\eta$
subset do not coincide. To diagonalize the mass
matrix and consequently the two-point functions
at \emph{leading order} we perform the following
transformation and find the corresponding \emph{lowest order} mixing angle    
\ba
\pi_{3} &=& \pi \cos(\epsilon) - \eta  \sin(\epsilon)   
\nonumber\\
\eta_{8} &=& \pi \sin(\epsilon) + \eta \cos(\epsilon) 
\ea
The lowest order mixing angle is 
\ba
\tan(2\epsilon) &=& \frac{\sqrt{3}}{2}\frac{m_d-m_u}{m_s-\hat m}\,,
\nonumber\\ 
\hat m &=& (m_u+m_d)/2\,.
\ea
G-parity requires the amplitude to vanish at the 
limit $m_u = m_d$ and therefore it must inevitably
be accompanied by an overall factor of $m_u-m_d$
which we have chosen to be in the form of $\sin(\epsilon)$.
\be
\label{defM}
A(\eta\to\pi^+\pi^-\pi^0) =
\sin(\epsilon)\, M(s,t,u)
\ee
Since the amplitude is invariant under charge 
conjugation we have 
\be
M(s,t,u) = M(s,u,t).
\ee
Note that the isospin breaking factor which is pulled out is
different in different references, \cite{GL3} and \cite{LeutwylerQM}
use different quantities. We have chosen $\sin(\epsilon)$ since it is the
factor that naturally shows up at lowest order.

For the eta decay to three neutral pions, 
the amplitude must be symmetric under
the exchange of pions (Bose symmetry) and this together with isospin symmetry
and using the fact that the decay is caused by the $\Delta I=1$
operator $(1/2) (m_u-m_d)\left(\bar u u-\bar d d\right)$ 
implies
\be
A(\eta\to\pi^0\pi^0\pi^0) = \sin(\epsilon)\,  {\overline M(s,t,u) } 
\ee
with
\be
\label{isorel}
{\overline M(s,t,u)} = M(s,t,u) +  M(t,u,s) + M(u,s,t)\,.
\ee
This relation is only true when isospin breaking in the amplitude
is taken into account to first order only. $M(s,t,u)$ and $\overline M(s,t,u)$
are treated in the isospin limit. We will work in this limit in the remainder
of the paper.

\subsection{A simplified form for the amplitude to order $p^6$}

The scattering amplitude can be represented
in terms of components with definite isospin assignments as\cite{Anisovich1}
\be
\label{defMi}
M(s,t,u) = M_0(s)+(s-u)M_1(t)+(s-t)M_1(u)+M_2(t)+M_2(u)-\frac{2}{3}M_2(s)\,.
\ee
The function $M_{I}(x=s,t,u)$ indicates scattering in the kinematic $x$-channel 
with total isospin I. 
Analyticity, unitarity and crossing symmetry
as imposed on the S-matrix, give rise to this exceptionally 
useful representation. For the derivation and detailed discussion 
for the case of $\pi \pi$ we refer to \cite{Knechtpipi}.  
This relation only holds up to ${\cal O} (p^8)$. 
The argument is based on the fact that nonpolynomial dependence
on $s,t,u$ is related to an absorptive part via unitarity.
Up to order $p^6$ there are only absorptive parts in the two-pion
$S$ and $P$-waves and then using isospin one derives
the form (\ref{defMi}).

It is important to note that the polynomial part of the 
amplitude cannot be split uniquely into $M_I$ functions
since the relation $s+t+u = 3 s_0$ allows a different redistribution
of the said part to the $M_I$s. A list of the equivalent
redefinitions for the case of $K\to3\pi$ can be found in App.~A
of \cite{BDP}. The choice we made is to remove as much as
possible first out of $M_2$ and then out of $M_1$.

Eq. (\ref{defMi}) makes the formidable task of handling two loop 
expressions much more manageable and we indeed confirmed 
explicitly the validity of Eq.~(\ref{defMi}) for the 
decay $\eta \to \pi^{o} \pi^{+} \pi^{-}$ at ${\cal O} (p^6)$. 

Note that the neutral decay amplitude can also be expressed
directly in terms of the $M_I(t)$,
\be
\overline M(s,t,u) = M_0(s)+M_0(t)+M_0(u)+\frac{4}{3}\left(
M_2(s)+M_2(t)+M_2(u)\right)\,.
\ee
when using the isospin relation (\ref{isorel}).

\subsection{Feynman Graphs}

We have collected all the amputated Feynman diagrams needed.
In these figures a filled circle denotes a vertex of order $p^2$, a filled
square a vertex of order $p^4$ and the grey filled square a vertex of order
$p^6$.
\FIGURE[t]{
\includegraphics[width=0.7\textwidth]{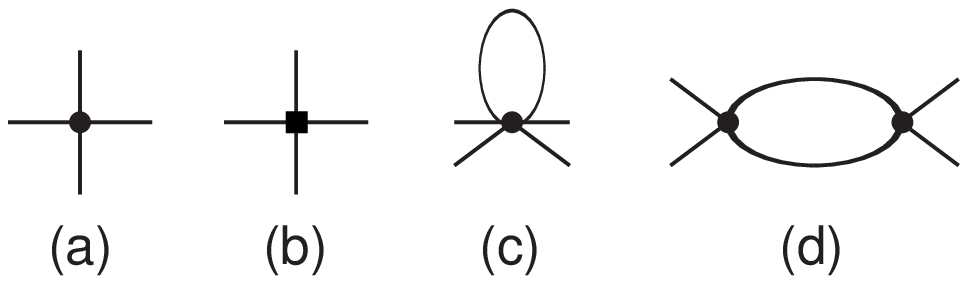}
\caption{
The Feynman diagrams of order $p^2$, (a) and of order $p^4$, (b)-(d).}
\label{figdiagrams1}
} 

\FIGURE[t]{
\includegraphics[width=0.65\textwidth]{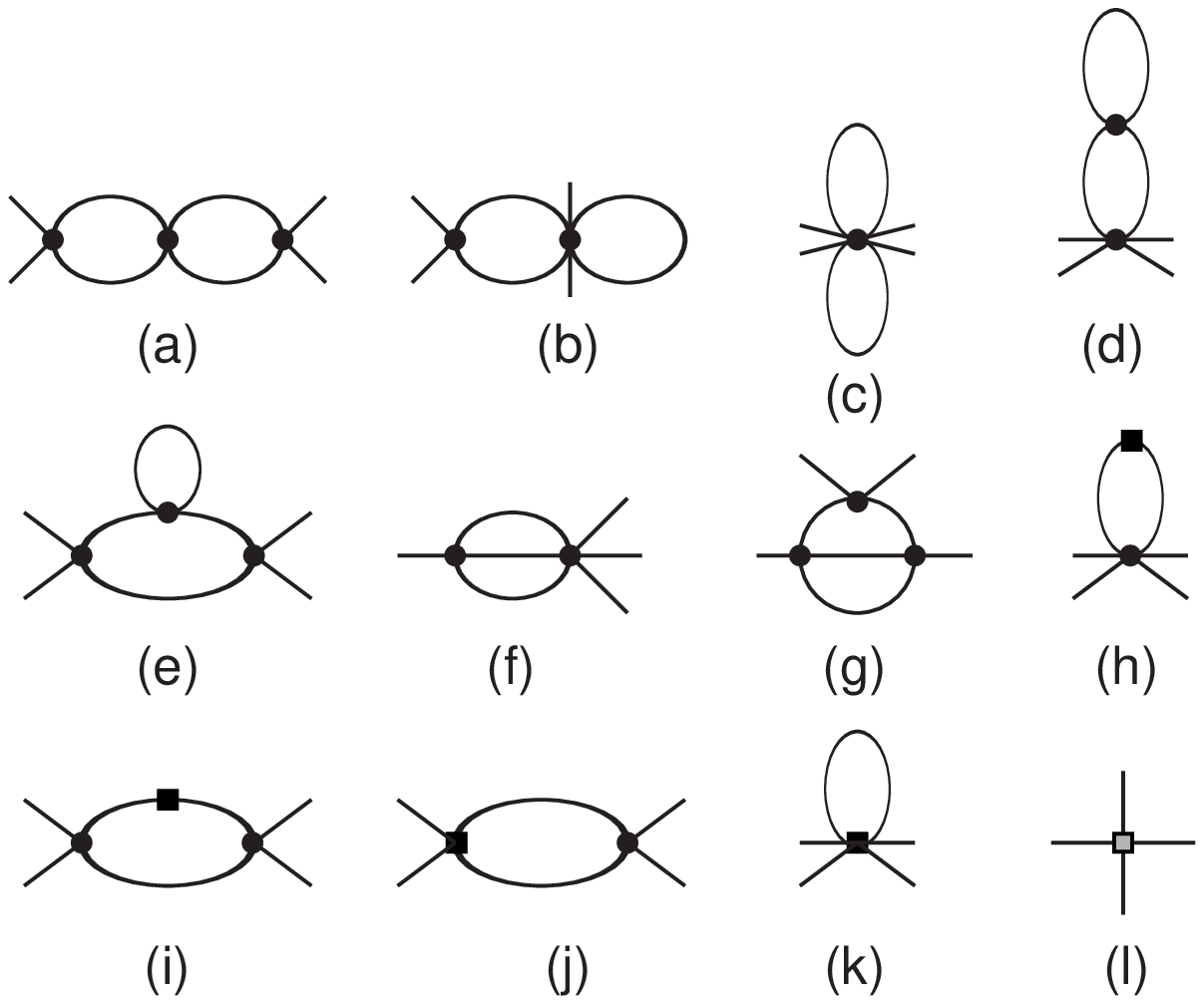}
\caption{
Feynman diagrams of order $p^6$.}
\label{figdiagrams2}
} 

\subsection{Regularization, renormalization and integrals}

As a regularization method we use dimensional integration. The regularization
method is described in detail in \cite{BCEGS2,BCE2}.

The one-loop functions needed are defined in many places. 
We use the definitions as given in \cite{ABT1,ABT3,BT2}.
The two-loop integrals we evaluate numerically using the methods described
in \cite{ABT1} for the sunsetintegrals and in \cite{BT2} for the
vertexintegrals.

In addition to the integrals given there, we also need derivatives w.r.t.
one of the masses in order to be able to pull out the overall isospin breaking
factor in the amplitudes. For all the one-loop integrals needed, these 
derivatives can be derived exactly. An example of a relation we derived,
needed to obtain agreement with the order $p^4$ result of \cite{GL3}
is
\ba
\overline C(m^2,m^2,m^2,p^2) &=& \left.\frac{\partial}{\partial m_1^2}
\overline B(m_1^2,m_2^2,p^2)\right|_{m_1^2=m_2^2=m^2}
\nonumber\\
&=& -\frac{2}{p^2-4 m^2}
\left(\overline B(m^2,m^2,p^2)-\overline B(m^2,m^2,0)-\frac{2}{16\pi^2}
\right)\,.
\ea
For the two-loop integrals, we have taken the derivatives numerically.

\section{Analytical results}
\label{sect:analytical}

\subsection{Order $p^2$}

At leading order there is one tree graph from ${\cal L}_2$ to compute
since we have already diagonalized the fields in the lowest order Lagrangian.
The lowest order decay amplitude takes on the form 
\ba
\label{resultp2}
M^{(2)}(s) = \frac{1}{F_\pi^2} \, \Big( \frac{4}{3}\,m_{\pi}^2 - s \Big)  
\ea
using the identity  $s+t+u = 3 s_0$. 
$F_\pi$ and $m_\pi$ are the physical pion decay constant and 
the physical pion mass respectively, which involve corrections of 
order $p^4$ and $p^6$. The higher order corrections due to the 
redefinition of the parameters are carried to the respective
amplitude of order $p^4$ and $p^6$. This agrees with the known lowest order
results \cite{orderp2x1,orderp2x2,GL3} but is written in a somewhat different
form. We have chosen this form since it brings out the Adler zero explicitly.
     
\subsection{Order $p^4$}

At this order we obtain vertices from ${\cal L}_2$ 
to construct the tadpole, Fig.~\ref{figdiagrams1}(c)
and the so-called unitarity graphs Fig.~\ref{figdiagrams1}(d)
and in addition, the tree diagram from ${\cal L}_4$,
Fig.~\ref{figdiagrams1}(b). These are the diagrams contributing to
${\cal G}_{1238}^{(4)}$ of (\ref{fullp4}).
The second set of terms in ${\cal A}_{1238}^{(4)}$ in (\ref{fullp4}) is from
what is usually called wave function renormalization.
The final two terms in  ${\cal A}_{1238}^{(4)}$ of (\ref{fullamplitude})
are what are called the mixing corrections\footnote{In our calculations,
lowest order mixing is dealt with exactly. The mixing at higher orders
is dealt with perturbatively.}.

The sum of all the contributions described in Eq.~(\ref{fullp4})
gives the full 
one-loop amplitude. G-parity requires the amplitude to be 
proportional to $\sin(\epsilon)$ as was the case at leading order.
This does not occur explicitly at order $p^4$, because now 
isospin is broken in the meson loops.     
In order to pull out an overall mixing angle we then 
carry out a Taylor expansion of the loop integrals involving 
two charged kaons around the neutral kaon mass. 
The amplitude is written as
\ba
\label{defMi4}
M^{(4)}(s,t,u) &=&\frac{1}{F_\pi^4}\bigg[ M^{(4)}_0(s)
+(s-u)M^{(4)}_1(t)+(s-t)M^{(4)}_1(u)
+M^{(4)}_2(t)+M^{(4)}_2(u)
\nonumber\\&&
-\frac{2}{3}M^{(4)}_2(s)\bigg]\,.
\ea
The full expression at ${\cal O}(p^4)$ for the $M^{(4)}(t)$
can be found in App.~\ref{App:p4}.
There, all the masses and the pion decay constant are the physical ones.
We have taken the expressions at lowest order and one-loop order
and added the correction terms needed at the higher orders to bring them into
the form we show in (\ref{resultp2}) and App.~\ref{App:p4}. 
and are corrected up to order $p^6$. We retain all the $p^4$ effective 
constants in our expression, in contrast to \cite{GL3} 
where all but $L_3$ are eliminated in favor of measurable quantities.  
We have also found our analytical results in full accord with that in 
\cite{GL3}.

\subsection{Order $p^6$}

We again split the amplitude as in (\ref{defMi}).
\ba
\label{defMi6}
M^{(6)}(s,t,u) &=&\frac{1}{F_\pi^6}\bigg[ M^{(6)}_0(s)
+(s-u)M^{(6)}_1(t)+(s-t)M^{(6)}_1(u)
+M^{(6)}_2(t)+M^{(6)}_2(u)
\nonumber\\&&
-\frac{2}{3}M^{(6)}_2(s)\bigg]\,.
\ea
However, the order $p^6$ expression is extremely long. The dependence on the
order $p^6$ LECs is given in App.~\ref{App:p6}.
We split the result in several parts
\be
\label{defMi6parts}
M^{(6)}_I(t) = M^C_I(t)+M^{LL}_I(t)+M^T_I(t)\,.
\ee
$M^C_I(t)$ contains the contributions from the order $p^6$ LECs,
$M^{LL}_I(t)$ contains the contributions that involve the order $p^4$ LECs
and $M^T_I(t)$ is the pure two-loop contribution, only dependent on the
masses of the pseudoscalars. $M^T_I(t)$ itself we split in the parts coming from
vertex-integrals, sunset-integrals and the rest. The latter split is
definitely not unique. It depends on how many of the relations between the
various integrals are actually used and we observe strong numerical
cancellations between its different parts.
We therefore only quote numerical results
for $M^T_I(t)$ as a whole.

The calculation has been performed independently by each of the authors,
the divergences agree with those of the general calculation \cite{BCE2}
and nonlocal divergences cancel as required. We have also checked explicitly
that the amplitudes can be brought into the form (\ref{defMi}).
The numerical results are also done twice and we have checked that they
agree. Finally, $\mu$-independence has been checked numerically and found
to be satisfactory. The latter is not exact, since we have expressed the
order $p^4$ and $p^6$ in the physical masses and there is a residual
$p^8$ $\mu$ dependence left over. We do observe a strong cancellation
between the order $p^4$ and $p^6$ $\mu$-dependence as expected.

\section{Resonance estimates of the $p^6$ LECs
and the inclusion of the $\eta^ \prime$ }
\label{sect:reso}

Chiral symmetry imposes no constraints on the values of the low energy 
constants, nevertheless these constants do depend on the parameters 
of the underlying theory, QCD, namely the masses of the heavy quarks and the 
 scale $\Lambda_{QCD}$. Hence, all the LECs may 
be determined from first principles
employing the Lattice QCD technique.             
At order $p^4$, most of the LECs have been determined phenomenologically and 
some of them are checked numerically by the use of the Lattice QCD
\cite{Necco}.
At the present time, only very few of the order $p^6$ LECs are estimated 
using available data\cite{reviewp6}.    
In this section, we will discuss briefly how to do an approximate estimation
of 
the LECs within the framework of the resonance effective field theories.   
In the limit $N_{c} \to \infty $, the matching of the QCD
and ChPT might become feasible since now, an infinite tower of massive
narrow hadronic states emerge. With these states one can
construct a chiral invariant Lagrangian
incorporating both Goldstone mesons and resonance fields.  
In practice one has to do a truncation on the hadronic 
spectrum and limit the resonance multiplets to low-lying excitations. 
The construction of the $p^4$ resonance Lagrangian is discussed in the 
pioneering works\cite{EGPR,EGLPR}.
The extention of the earlier works 
to ${\cal O} (p^6)$ is presented in \cite{Cirigliano1}.
In the following we present a resonance Lagrangian at order $p^6$
with the vector realization of the vector fields, \cite{EGPR,Cirigliano1}
use the antisymmetric tensor formulation.
The exchange of axialvector resonances does not contribute
to the process $\eta \to 3\pi$ and they are not discussed here.
We do, in addition to the lowest vector nonet, include the pseudoscalar singlet
and a scalar nonet.

The building blocks one needs for the construction of the Lagrangian
take the form 
\ba 
u^\mu &=& i u^\dagger D^\mu U u^\dagger\quad\mathrm{with}\quad u^2 = U\,,
\nonumber\\
\chi_{\pm} &=& u^\dagger \chi u^\dagger \pm  u \chi^\dagger  u\,.
\hspace{-1cm}
\ea
These transform as an octet under $SU(3)_{V}$ using the general
CCWZ construction\footnote{See Refs. \cite{EGPR,Cirigliano1} for a
more extensive discussion and references}.

The resonance fields are counted as order 1 in the chiral counting.
We do not include the complete possible list of terms here but restrict
to a smaller subset which contains the lowest order interactions
of the resonances in our chosen representation for them.
This is the same subset of possible terms used in most of the work
on NNLO ChPT. We only show the terms relevant for $\eta\to3\pi$ here.
For the vector fields we use as Lagrangian
\be
\label{lagV}
{\cal L}_V = -\frac{1}{4}\langle V_{\mu\nu}V^{\mu\nu}\rangle
+\frac{1}{2}m_V^2\langle V_\mu V^\mu\rangle
-\frac{ig_V}{2\sqrt{2}}\langle V_{\mu\nu}[u^\mu,u^\nu]\rangle
+f_\chi\langle V_\mu[u^\mu,\chi_-]\rangle\,.
\label{vector}
\ee
$V_{\mu\nu} = \nabla_{\mu} V_{\nu} - \nabla_{\nu} V_{\mu}$. 
The singlet component does not contribute to order $p^6$
for $\eta\to3\pi$.
For the scalar meson nonet, the matrix of fields $S$, we consider
\be
\label{lagS}
{\cal L}_S = \frac{1}{2} \langle \nabla^\mu S \nabla_\mu S 
 - M^2_S S^2 \rangle  
 + c_d \langle Su^\mu u_\mu \rangle + c_m \langle S \chi_+ \rangle 
\label{scalar}
\ee
At tree level integrating out the heavy fields is equivalent to 
applying the equation of motion for the elimination of the heavy 
fields. To solve the equation of motion we perform a perturbative 
expansion of the resonance field 
with coefficients in increasing powers of
$1/M_{R}$ ($M_{R}$, the resonance mass) 
and then solve the equation of motion recursively.   
We obtain for the Vector $V^\mu$ and Scalar $S$ resonance field
\ba
\label{solVS}
V^\mu &=& -\frac{ig_V}{\sqrt{2}M^2_V} \nabla_\nu[u^\nu,u^\mu] 
-\frac{1}{M^2_V}f_\chi [u^\mu,\chi_-]+\cdots
\nonumber\\
S &=&  \frac{c_d}{M^2_S} (u_\mu u^\mu) 
+ \frac{c_m}{M^2_S}(\chi_+) +\frac{c_d}{2M^4_S} 
 \nabla^\nu \nabla_\nu (u_\mu u^\mu) 
+ \frac{c_m}{2M^4_S} (\nabla^\mu \nabla_\mu \chi_+ )+\cdots\,.
\ea
Substitution of (\ref{solVS}) in the Lagrangians (\ref{lagV}) and (\ref{lagS})
we obtain the effective action from $V,S$-exchange\cite{BCEGS2,ABT3}
\ba
\label{LagInt}
{\cal L}_V &=& -\frac{1}{4 M^2_V} \left\langle \left( i g_V\,
\nabla_\mu [ u^\nu,u^\mu ] - f_\chi \sqrt{2}\, [ u^\nu,\chi_- ] \right)^2 
\right\rangle
\\
\label{LagInt2}
{\cal L}_S &=& \frac{1}{2 M^4_S} \left\langle 
\left( c_d \nabla^\nu ( u_\mu u^\mu )
 + c_m \nabla^\nu \chi_+ 
 \right)^2 \right\rangle
\ea
The resonance couplings were determined in \cite{BCEGS2,ABT3,EGPR}.
The values we use are
\be
 f_\chi = -0.025,\quad  g_V = 0.09,\quad 
 c_m = 42 \mbox{ MeV},\quad
  c_d = 32 \mbox{ MeV}, 
\ee
and for the masses we use
\be
m_V = m_\rho = 0.77 \mbox{ GeV}, 
m_S = 0.98 \mbox{ GeV}.
\ee

The $\eta^ \prime$ plays a significant role in processes 
involving $\eta$ due to the $\eta-\eta^ \prime$ mixing.
Within the quark model, $\eta^ \prime$ and $\eta$ mix 
because of the $SU(3)$ symmetry breaking. 
In the chiral limit the pseudoscalar octet becomes 
massless but the $\eta^ \prime$ has a residual mass as a 
result of the axial $U(1)$ anomaly. In the combined
chiral and large $N_{c}$ limit, however, a nonet of Goldstone 
particles emerge and this provides a perturbative framework 
to investigate the dynamical interplay between $\eta^ \prime$
and Goldstone particles, see e.g. \cite{LeutwylerEEP,KL}.

One important result is the 
saturation of the low energy constant $L_{7}$ by
the $\eta^ \prime$ exchange\cite{GL1}
\be
\label{valL7}
L^{\eta \prime}_{7}  =  -\frac{\tilde d_m^2}{2 M^2_{\eta \prime}}
\ee  
We therefore perceive the $\eta^ \prime$ dynamical 
effects at order $p^4$ through its contribution to the 
effective constant $L_7$. In the light of this 
realization, the $\eta-\eta^ \prime$ mixing effect on the $C_{i}$ involved
in the decay $\eta \to 3\pi$ can be obtained by constructing 
an appropriate $U(3)$ Lagrangian at order $p^6$.
That  $\eta-\eta^ \prime$  mixing can be treated perturbatively
for the decay $\eta \to 3\pi$ is discussed extensively
in \cite{LeutwylerEEP}.   
We take for the singlet degree of freedom $P_1$ the simple Lagrangian     
\be
\label{lageta}
{\cal L}_{\eta^\prime} = \frac{1}{2}\partial_\mu P_1\partial^\mu P_1
-\frac{1}{2}M_{\eta^\prime}^2 P_1^2
+i\tilde d_m P_1\langle\chi_- \rangle\,.
\ee
Integrating out $P_1$ leads to the order $p^4$ term with $L_7$
of (\ref{valL7}) and the order $p^6$ Lagrangian
\be
\label{etaprime}
{\cal L}_{\eta'} =  
- \frac{\tilde{d}_m^2}{2 M^4_{\eta'}} 
\partial_\mu \langle \chi_- \rangle \partial^\mu \langle \chi_- \rangle
~\mbox{with}~ \tilde{d}_m = 20 \, \, \mbox{MeV}. 
\ee
The latter can be rewritten in general in terms of the basis of operators
of \cite{BCE1}. The result is\footnote{This was derived by the authors of
 \cite{ABT3} but not included in the final manuscript.
It agrees with the expression shown by Kaiser\cite{Kaiser}.}
\ba
\label{relOi}
\partial_\mu\langle\chi_-\rangle\partial_\mu\langle\chi_-\rangle
&=& O_{18}+\frac{2}{9}O_{19}-\frac{1}{3}O_{20}+\frac{1}{3}O_{21}
+ 2 O_{27}+\frac{2}{3}O_{31}-\frac{1}{3}O_{32}+\frac{1}{3}O_{33}
\nonumber\\&&
-2O_{35}
+ O_{37}-\frac{8}{3}O_{94}\,.
\ea
\TABLE[t]{
  \centering
  \begin{tabular}{|c|ccccccc|c|}
  \hline
  i & $\frac{ F_0^2 g_V^2}{M_V^2}$ & $\frac{F_0^2 g_V f_\chi}{\sqrt2 M_V^2}$ &
   $\frac{F_0^2 f_\chi^2}{M_V^2}$ &
   $\frac{F_0^2c_d^2}{M_S^4}$ & $\frac{F_0^2 c_d c_m}{M_S^4}$ &
  $\frac{F_0^2 c_m^2}{M_S^4}$ &  $\frac{F_0^2 \tilde d_m^2}{M_{P_1}^4}$\\
  \hline
1 &1/8 & & &$-$1/4 & & & \\  
4 &1/8 & & & & & & \\  
5 & & & & &1/2 & & \\  
8 & & & & &1/2 & & \\  
10 & & & & &$-$1 & & \\  
12 & & & & &$-$1/2 & & \\  
18 & & & & & & &$-$1/2 \\  
19 & & & & &1/27 & &$-$1/9 \\  
20 & & & & &$-$1/18 & &1/6 \\  
22 &1/16 &1/2 & &1/8 & & & \\  
24 &1/12 & & &$-$1/6 & & & \\  
25 &$-$3/8 &$-$1 & &1/4 & & & \\  
26 &7/36 & 1 & 1 &$-$5/36 &$-$1/2 &$-$1/4 & \\  
27 &$-$1/36 & & &1/18 &1/3 & &$-$1 \\  
28 &1/72 & & &$-$1/36 & & & \\  
29 &$-$11/72 &$-$1 & $-$1 &1/18 &$-$1/2 &$-$1/4 & \\  
31 & & & & &$-$7/18 & &$-$1/3 \\  
32 & & & & &$-$1/18 & &1/6 \\  
33 & & & & &2/9 & &$-$1/6 \\  
\hline
  \end{tabular}
  \caption{The resonance exchange estimates of $C_i$ contributing to
$\eta\to3\pi$. The vector exchange results are taken from \cite{Kampf}
scalar exchange from \cite{Cirigliano1}. For the singlet eta contribution,
see text. The result for the $C_i^r$ is the top row multiplied
by the coefficients given in the table. Only nonzero coefficients that also
contribute to $\eta\to3\pi$ are given. We use here the dimensionless
version of the $C_i^r$. Only terms relevant for $\eta\to3\pi$ are shown.
}
  \label{tabresonance}
} 

We have derived the contribution to $\eta\to3\pi$ in different ways.
An option is to evaluate explicitly the resonance exchange directly from
the Lagrangians with resonances. The second method is as described above,
to evaluate the exchange in general in terms of an effective Lagrangian of
the pseudoscalars only and then calculate with that one.
The third option is to rewrite the contributions from vector and resonance
exchange in terms of the order $p^6$ ChPT Lagrangian \cite{BCE2}.
The latter can be done for the scalar octet using the results of
\cite{Cirigliano1}, for the vector nonet using \cite{Kampf}
and the pseudoscalar singlet using (\ref{relOi}).
We have checked that the first method also gives the Lagrangians
(\ref{LagInt}) and (\ref{LagInt2}) and that the 2nd and third method
in the end give the same contribution to $\eta\to3\pi$.
The results for the $C_i$ are quoted in Tab.~\ref{tabresonance}.
In the numerical estimates we have used $F_0=F_\pi$.

\section{Numerical results}
\label{sect:numerical}

\subsection{A first look}
\label{sect:firstlook}

In this section we present some plots of the amplitude to give a first
impression of how the higher orders look like. We start with figures showing
different contributions to $M_{0,1,2}(s)$. It should be noted that since
terms can be moved around between the $M_I(s)$ these figures cannot be used
to check whether we have convergence or not. They are shown for
illustration only. We have actually plotted the negative of the quantities
defined earlier, this makes the size of the corrections easier to
judge.

The input values we use are the physical eta mass, $m_\eta=547.3$~MeV,
the average Kaon mass removing electromagnetic effects, $m_K = 494.53$~MeV
and a pion mass such that $s+t+u=m_\eta^2+3m_\pi^2$ is satisfied for the
charged and neutral decay. So we use $3m_\pi^2=2m_{\pi^+}^2+m_{\pi^0}^2$
for the charged decay and $m_\pi^2=m_{\pi^0}^2$ for the neutral decay.
More general plots were always done with the mass as for the charged decay.
The order $p^4$ LECs are set to the values for fit 10 of \cite{ABT4}
and we have set the order $p^6$ LECs equal to zero. The subtraction
scale is $\mu=770$~MeV.

\FIGURE[t]{
\begin{minipage}{0.48\textwidth}
\includegraphics[width=\textwidth]{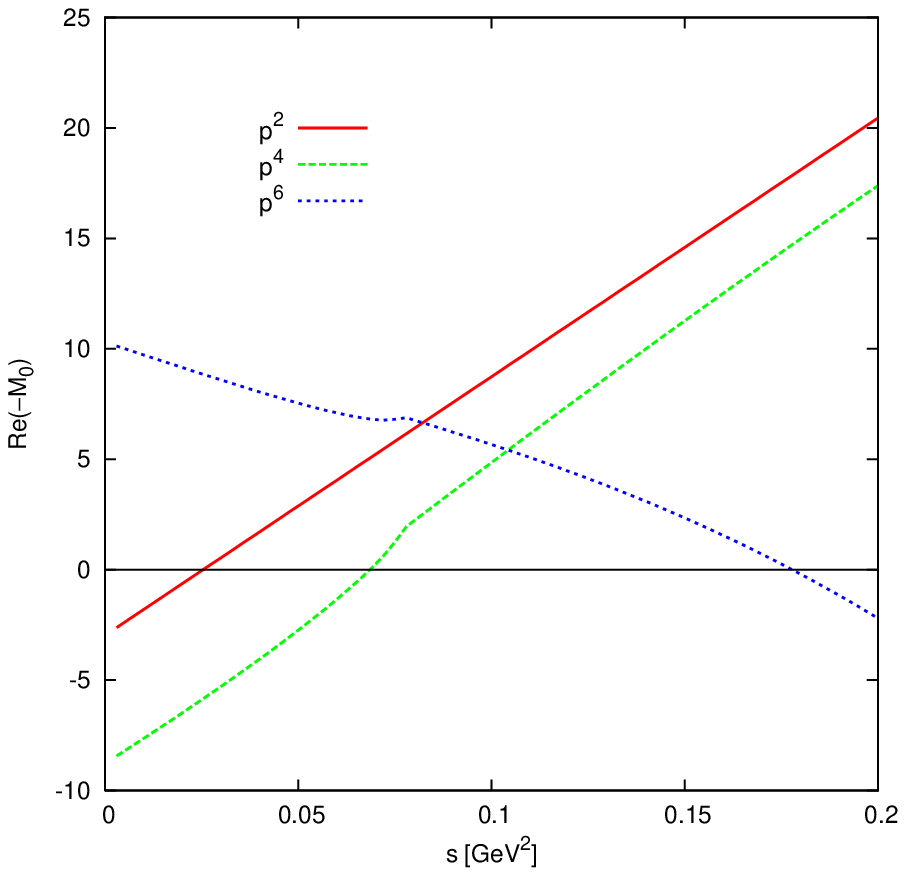}
\centerline{(a)}
\end{minipage}
\begin{minipage}{0.48\textwidth}
\includegraphics[width=\textwidth]{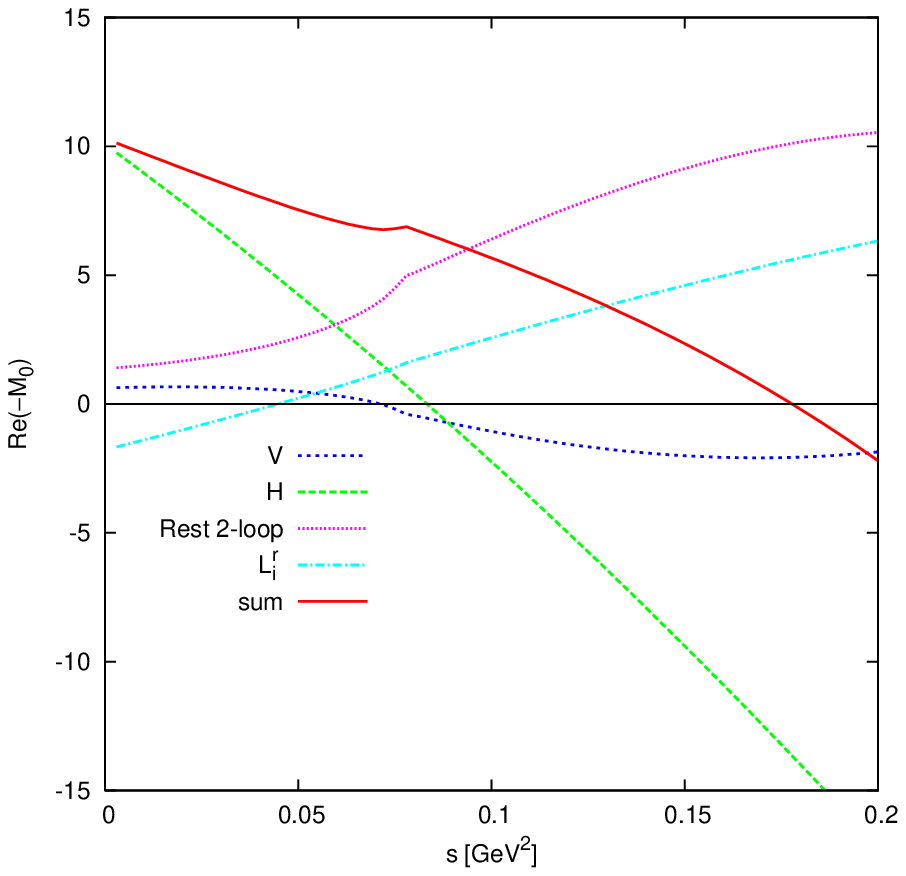}
\centerline{(b)}
\end{minipage}
\caption{
(a) The order $p^2$, $p^4$ and $p^6$ contribution to $-\mathrm{Re} M_0(s)$.
(b) The order $p^6$ contribution to $-M_0(t)$ split into its parts, the
contribution from vertex-integrals (V), sunsetintegrals (H)
and the remaining pure two-loop part as well as the $L_i^r$-dependent part.
}
\label{figReM0}
} 
 In Fig.~\ref{figReM0}(a) we show the contributions of order $p^2$,
$p^4$ and $p^6$ to $M_0(t)$. One sees an acceptable convergence in the
physical domain for the decay. In Fig.~\ref{figReM0}(b) we show
how the different parts contribute. As one can see, there are sizable
cancellations in the order $p^6$ contribution.
\FIGURE[t]{
\begin{minipage}{0.48\textwidth}
\includegraphics[width=\textwidth]{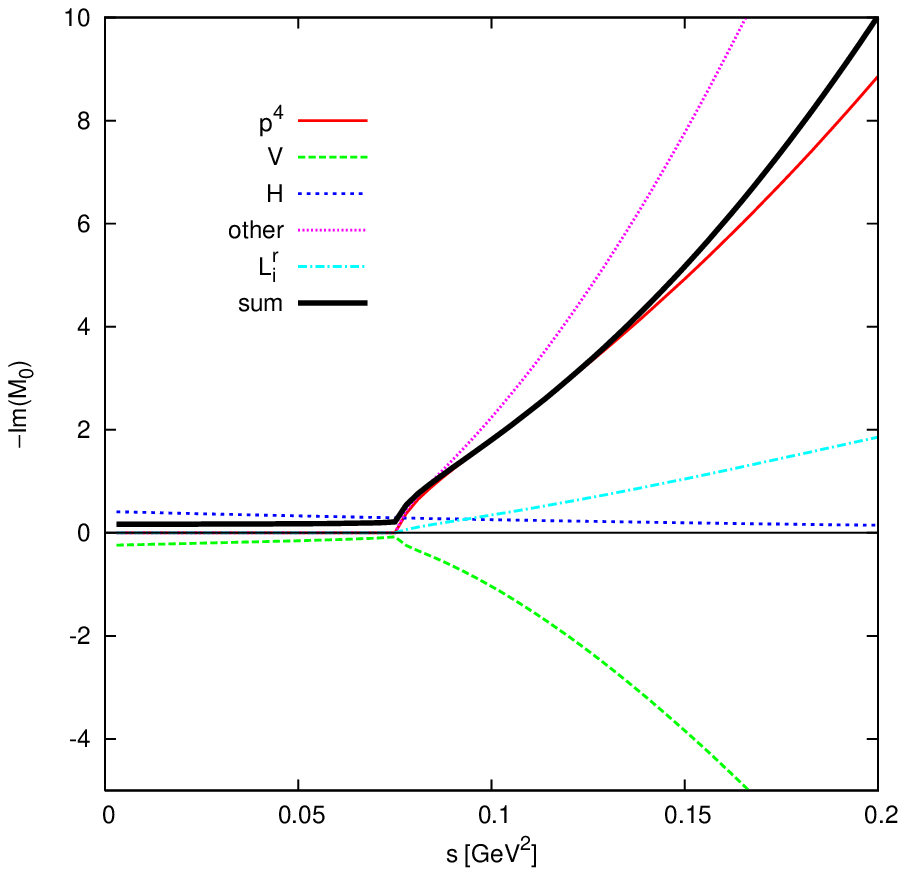}
\centerline{(a)}
\end{minipage}
\begin{minipage}{0.48\textwidth}
\includegraphics[width=\textwidth]{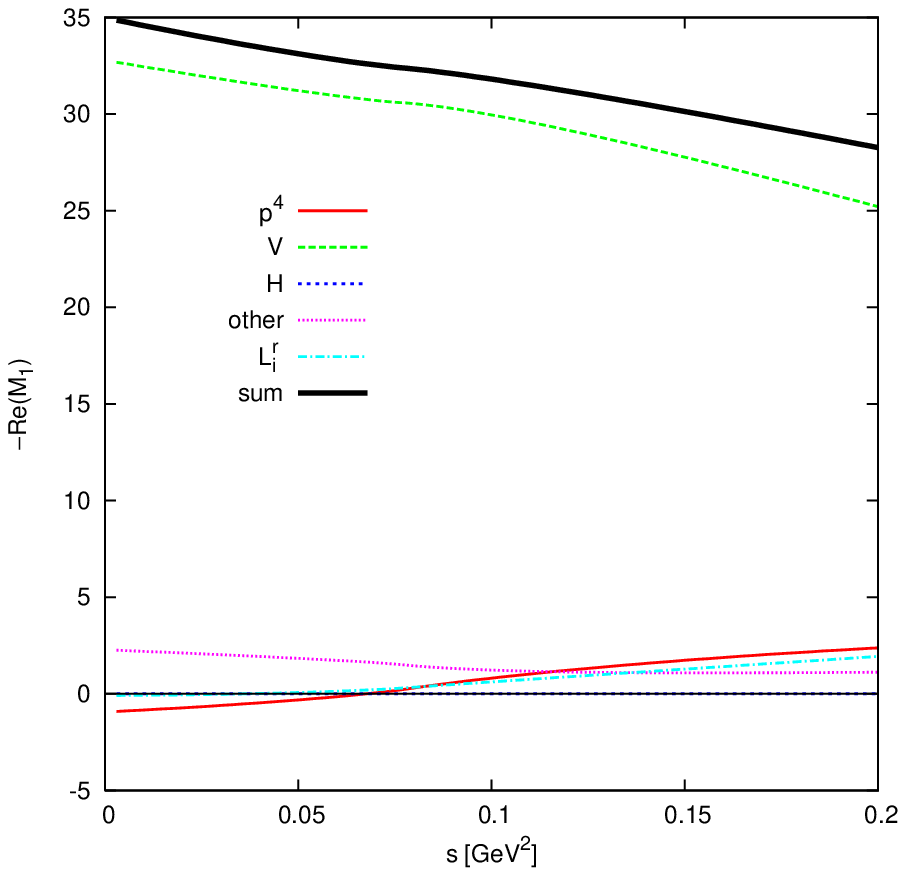}
\centerline{(b)}
\end{minipage}
\caption{
(a) The order $p^2$, $p^4$ and $p^6$ contribution to $-\mathrm{Im} M_0(s)$.
(b) The order $p^4$ and $p^6$ contribution to $-\mathrm{Re} M_1(s)$.
}
\label{figImM0}
} 
In Fig.~\ref{figImM0}(a) we show the various contribution to the
absorptive part of $M_0(s)$. We see that the total $p^6$ is about the
same size as the order $p^4$ one. To be noted is the three particle
cut that contributes first at order $p^6$. This allowed $M_0(s)$ to have
an absorptive part also below the two-pion threshold. 
 This cut gets contributions
from the vertex- and the sunset-integrals, 
diagrams in Fig.~\ref{figdiagrams2}(f) and (g).
As expected,
this cut gives a rather small contribution.
\FIGURE[t]{
\begin{minipage}{0.49\textwidth}
\includegraphics[width=\textwidth]{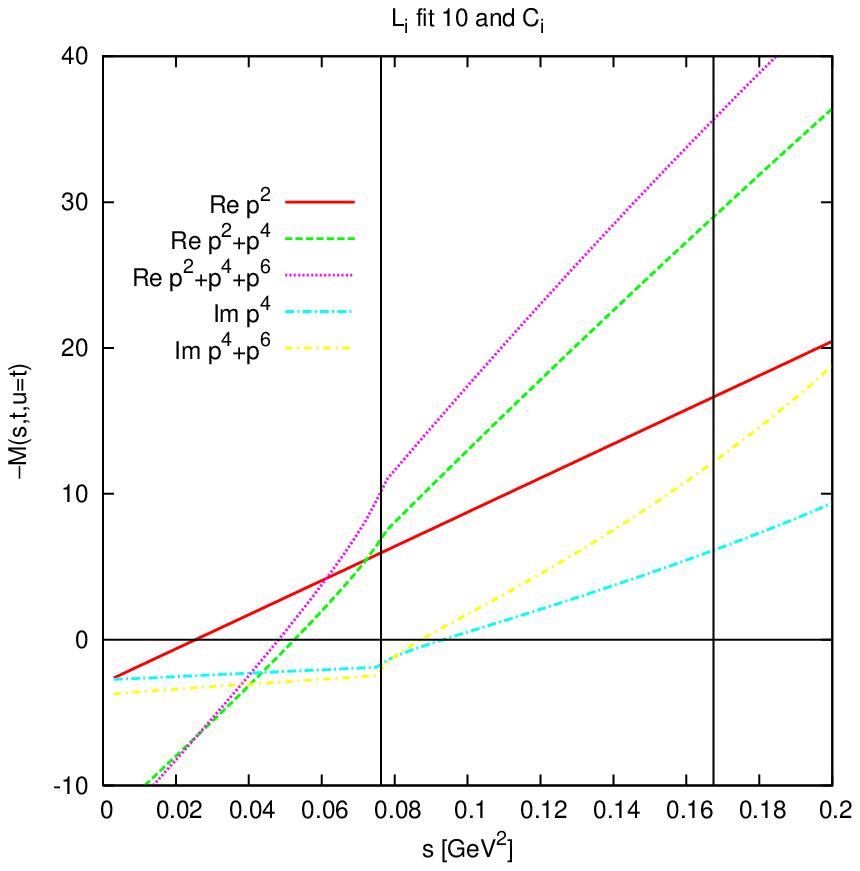}
\centerline{(a)}
\end{minipage}
\begin{minipage}{0.49\textwidth}
\includegraphics[width=\textwidth]{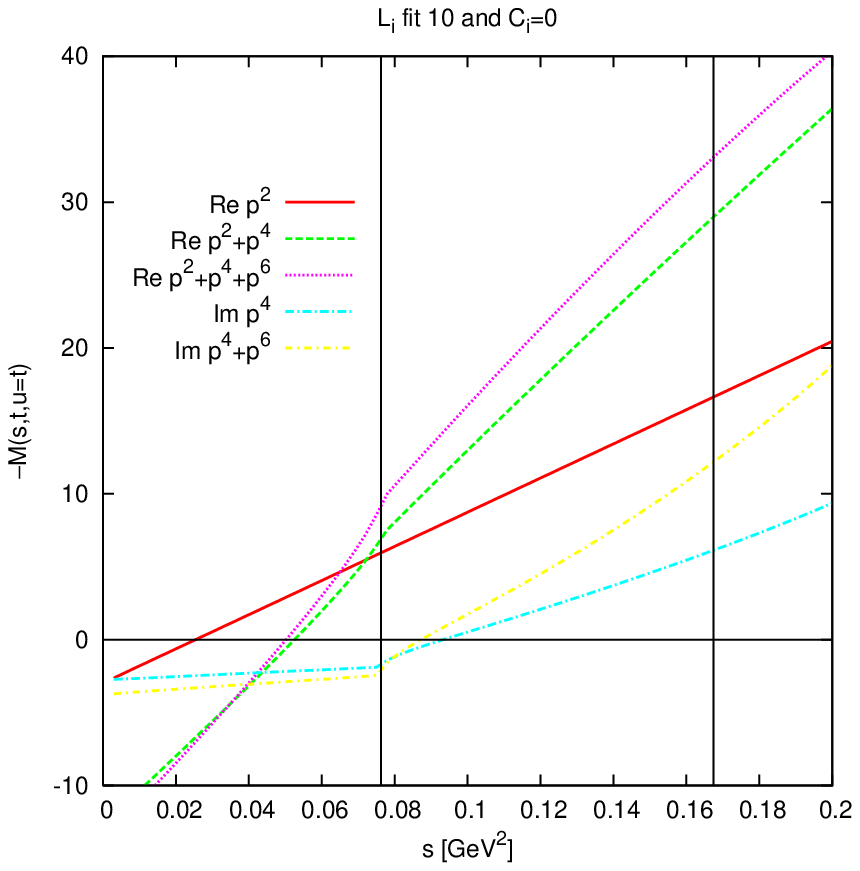}
\centerline{(b)}
\end{minipage}
\caption{The amplitude $M(s,t,u)$ along the line $t=u$. The vertical
lines indicate the physical region.
(a) Shown are the
real and imaginary parts with all parts summed up to the given order.
(b) Same plot but the contribution from the $C_i^r$ has been removed.}
\label{figMstu1}
} 
We also show a case where the convergence looks bad, 
$M_1(s)$ shown in Fig.~\ref{figImM0}(b). When we look at the full amplitude
this large correction is not visible. It seems to be tempered sufficiently
when all the different $M_I(s)$ are summed as in (\ref{defMi}).

\FIGURE[t]{
\begin{minipage}{0.49\textwidth}
\includegraphics[width=\textwidth]{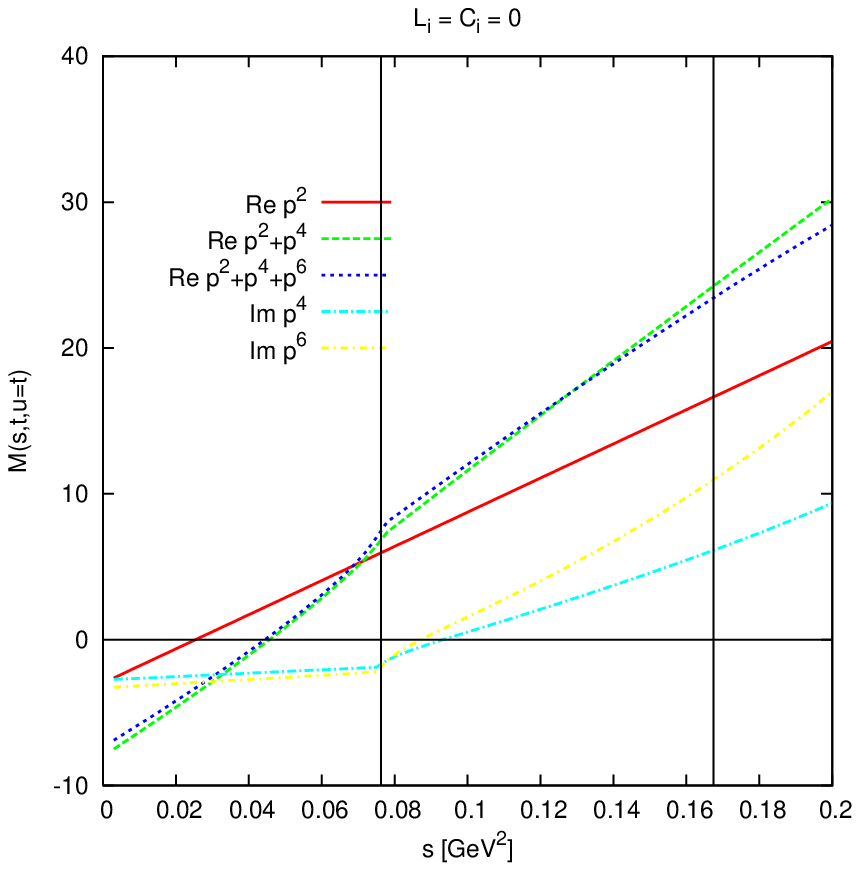}
\centerline{(a)}
\end{minipage}
\begin{minipage}{0.49\textwidth}
\includegraphics[width=\textwidth]{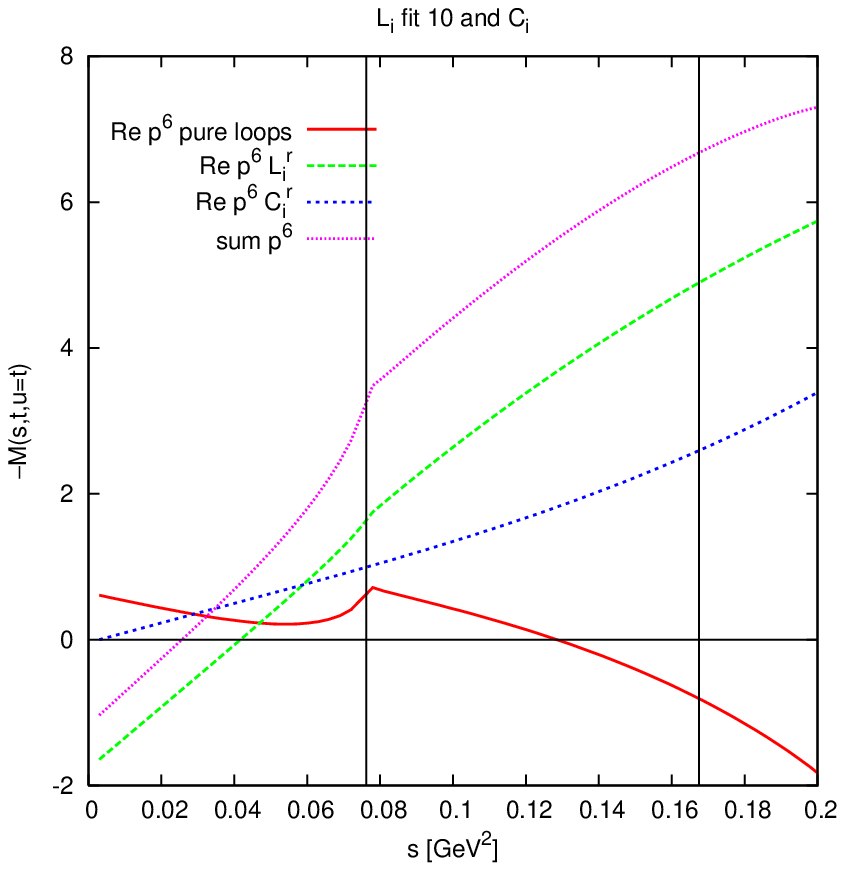}
\centerline{(b)}
\end{minipage}
\caption{The amplitude $M(s,t,u)$ along the line $t=u$.
The vertical lines indicate the physical region. 
(a) Shown are the
real and imaginary parts with all parts summed up to the given order
but the contribution from the $L_i^r$ and $C_i^r$ have been removed.
(b) Same plot but showing the various parts.}
\label{figMstu2}
} 

\FIGURE[!t]{
\begin{minipage}{0.49\textwidth}
\includegraphics[width=\textwidth]{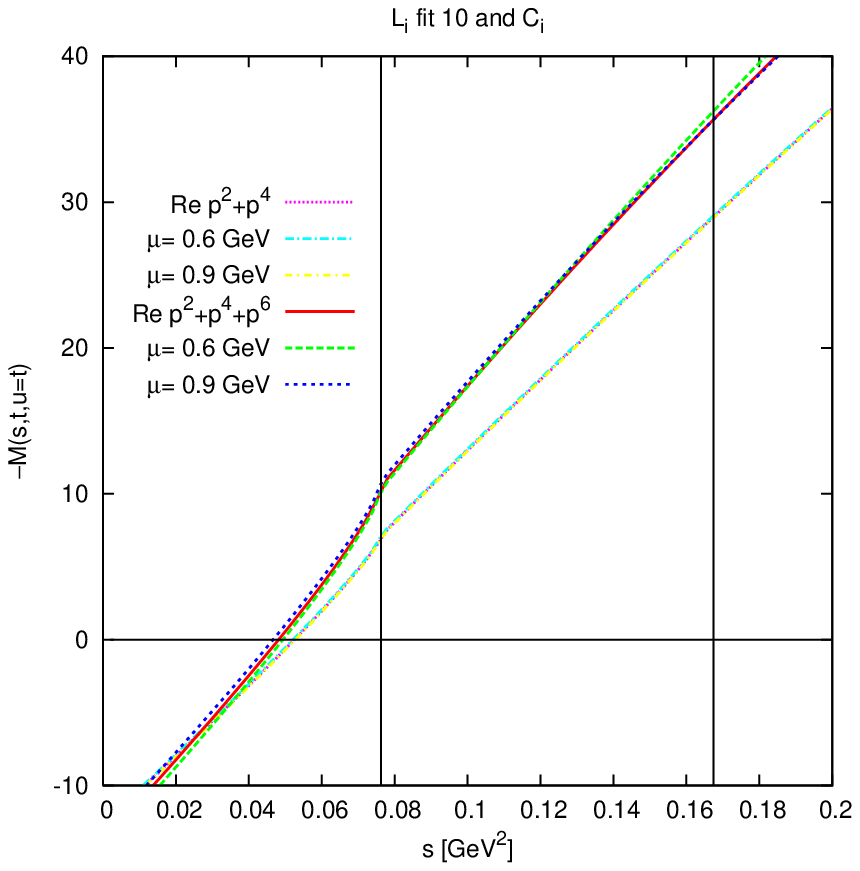}
\end{minipage}
\caption{The amplitude $M(s,t,u)$ along the line $t=u$
to NLO and NNLO order for three choices of the subtraction point $\mu$,
i.e. three choices at which the estimate of the $C_i^r(\mu)$ is applied.
}
\label{figMstumu}
} 
To see this, let us look at the amplitude along the line $u=t$ as a function
of $s$ for the full amplitude. Here we plot first our full result.
In Fig.~\ref{figMstu1} we show the order $p^2$, the sum of order $p^2$
and $p^4$ and finally the sum of order $p^2$, $p^4$ and order $p^6$.
Note the shift in the Adler zero in going from order $p^2$ to order $p^4$.

As a final part here we show the dependence on the subtraction constant $\mu$.
As mentioned above, our full result is $\mu$-independent to the order it
should be with a cancellation between the variation at NLO and NNLO.
The $\mu$-depen\-dence creeps in via the estimate of the $C_i^r$
and at which scale $\mu$ it is applied.
In Fig.~\ref{figMstumu} we plot the real part of the amplitude to NLO
and NNLO at $\mu=0.6$, $0.77$ and $0.9$~GeV. The variation
with $\mu$ is fairly small.

\subsection{Comparison with the dispersive result}
\label{sect:dispersive}

Since long ago it has been known that the decay amplitude of 
$\eta \to \pi \pi \pi$ receives a sizable enhancement due to 
the loop correction at ${\cal O} (p^4)$ \cite{GL3}.        
In fact it turned out that a large part of this correction
comes from $\pi\pi$ rescattering in the final state as
expected\cite{unitarity1,unitarity2}.
This has prompted two different analyses using dispersive methods.
They both restrict themselves to $\pi\pi$-rescattering but differ in how
the subtraction constants are determined and in how the dispersion theory
is used. Ref.~\cite{Kambor1} used the Khuri-Treiman equations
and fixed the subtraction constants by comparing with the one-loop expression
at various kinematical points. Ref.~\cite{Anisovich1} generalized
the reasoning behind \cite{Knechtpipi} to obtain a series of
dispersion relations for the $M_I(t)$ defined in (\ref{defMi}).
Their predictions on the decay width
are in agreement within the quoted uncertainties,
\cite{Anisovich1} finds $\Gamma = 219\pm22$~eV and \cite{Kambor1}
obtain $\Gamma = 209\pm20$~ eV. Both used
Dashen's theorem and the then known value of $L_3^r$ to fix the overall
constant.  
These, however, must be compared with $\Gamma_{exp} = 295$~eV which provides
a check on Dashen's theorem~\cite{LeutwylerQM}.
On the other hand, as emphasized in \cite{BG,BijnensETA05},
they have a rather different behaviour in the phasespace distributions.
This is shown in Fig.~\ref{figdispersive}. It can be seen that the slope
for \cite{Kambor1} is smaller than the order $p^4$ results while
\cite{Anisovich1} has a larger slope than the one-loop result.
The latter also follows from the very simplified dispersive analysis
performed in \cite{BG} shown in Fig.~\ref{figdispersive2}(a). The general
feature of the result of \cite{Anisovich1} seems to be robust against
small changes.
One motivation for the present work is to check these predictions and see
if other effects could be important as well, at order $p^4$ only about half
the correction came from the unitarity correction.

We can now compare the same plot coming from our full NNLO
calculation shown in Fig.~\ref{figdispersive2}(b).
The total order $p^6$ correction is somewhat larger than observed in
\cite{Kambor1,Anisovich1} but the trend is definitely in better agreement
with \cite{Anisovich1}.
\FIGURE[t]{
\begin{minipage}{0.48\textwidth}
\includegraphics[width=\textwidth]{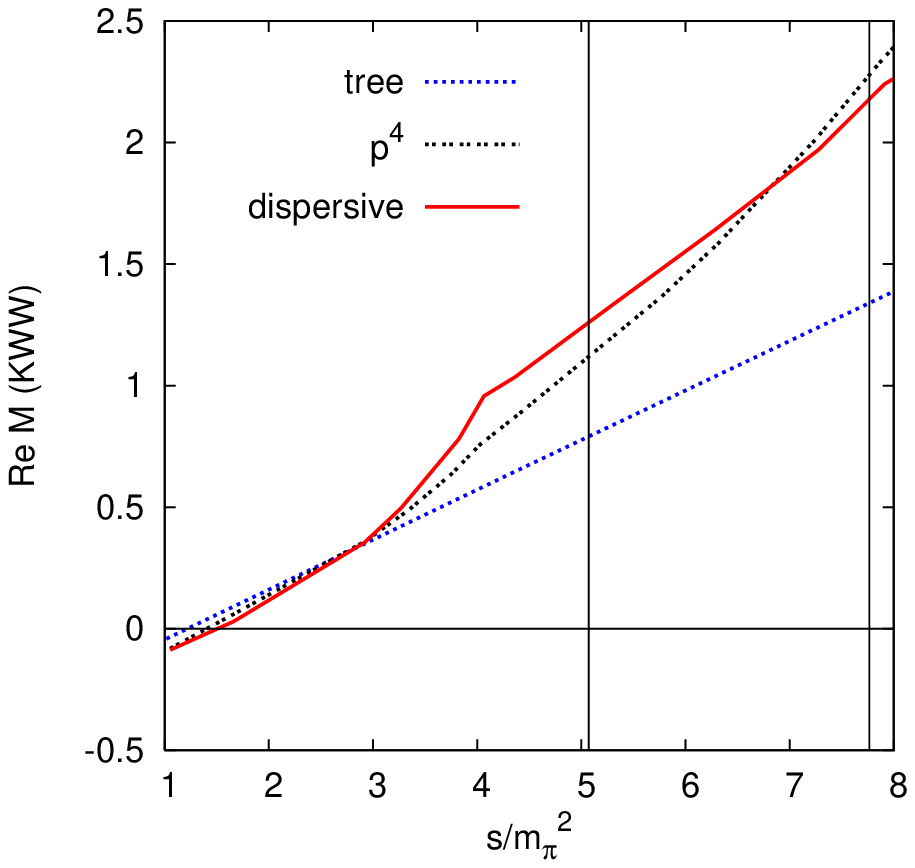}
\centerline{(a)}
\end{minipage}
\begin{minipage}{0.48\textwidth}
\includegraphics[width=\textwidth]{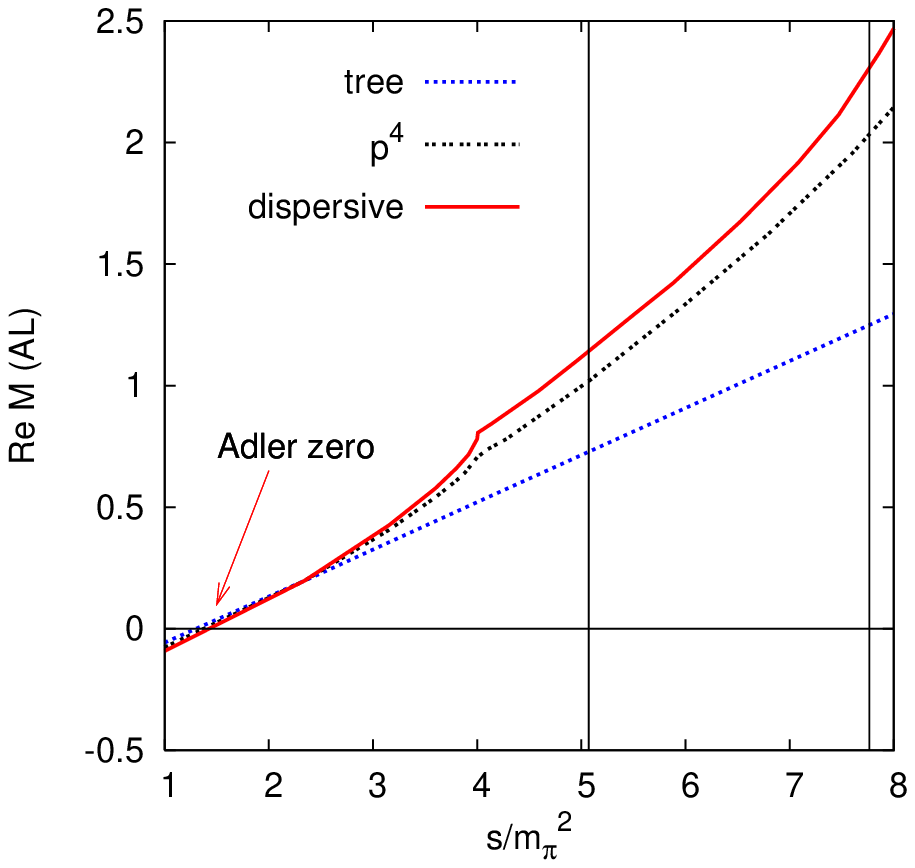}
\centerline{(b)}
\end{minipage}
\caption{
(a) Decay amplitude obtained by use of extended Khuri-Treiman
equations\cite{Kambor1} along the line $s=u$.
(b) Alternative dispersive analysis for the decay amplitude\cite{Anisovich1}.
Figs. from \cite{BijnensETA05}, adapted from \cite{Kambor1,Anisovich1}.
}
\label{figdispersive}
} 
\FIGURE{
\begin{minipage}{0.48\textwidth}
\includegraphics[width=\textwidth]{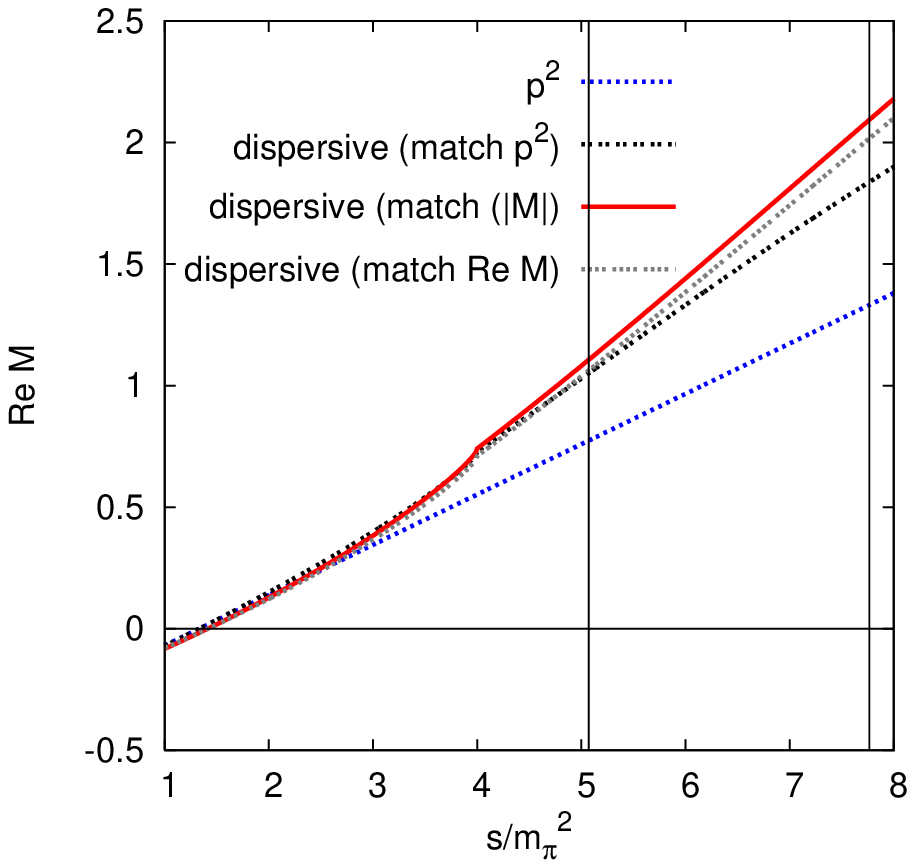}
\centerline{(a)}
\end{minipage}
\begin{minipage}{0.48\textwidth}
\includegraphics[width=\textwidth]{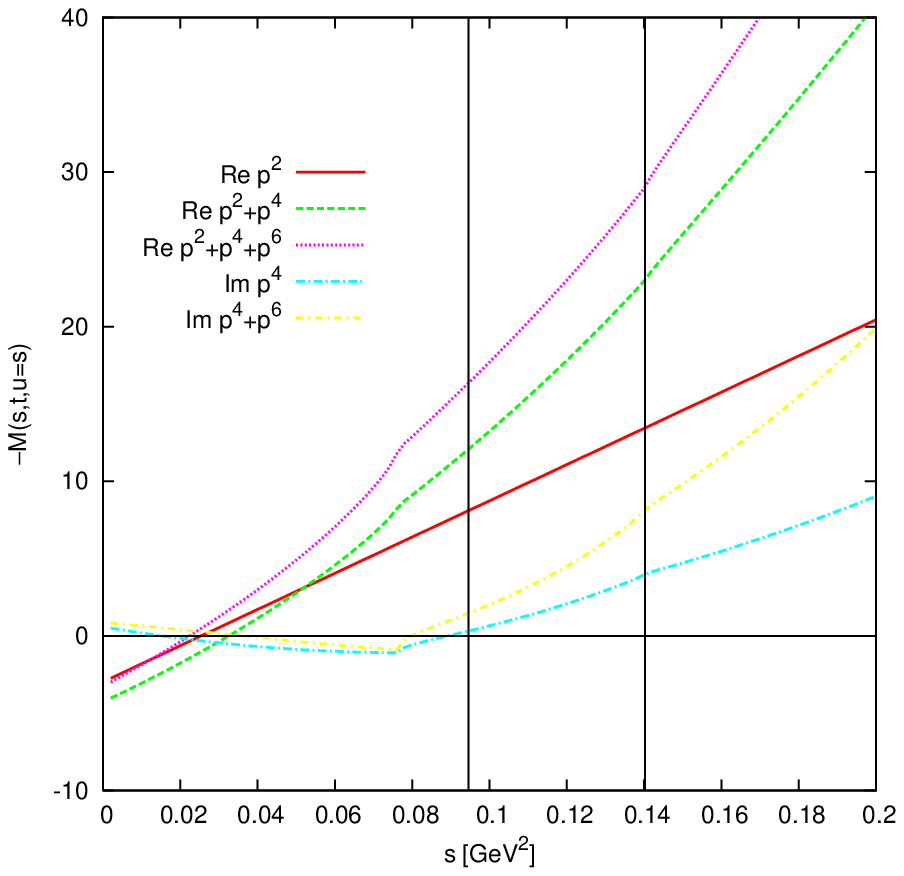}
\centerline{(b)}
\end{minipage}
\caption{
(a) 
The simplified analysis of \cite{BG} which shows the same general behaviour
as the result of \cite{Anisovich1} shown in Fig.~\ref{figdispersive}(b).
(b) Our result for $M(s,t,u)$ at $s=u$.}
\label{figdispersive2}
} 

In \cite{Anisovich1} the fact that the position of the Adler zero
did not change much from LO to NLO was used to determine the
subtraction constants. In Fig.~\ref{figadler} we show a blowup
of the region around the Adler zero. As can be seen, the position
of the zero in the real part varies a bit by going from LO to NLO
and then almost moves back at NNLO. This is not incompatible
with the observation of \cite{Anisovich1}, we use a slightly different
version of NLO then the one in \cite{GL3} and use different input parameters
resulting from the order $p^6$ determination of LECs \cite{ABT4}.
\FIGURE{
\begin{minipage}{0.48\textwidth}
\includegraphics[width=\textwidth]{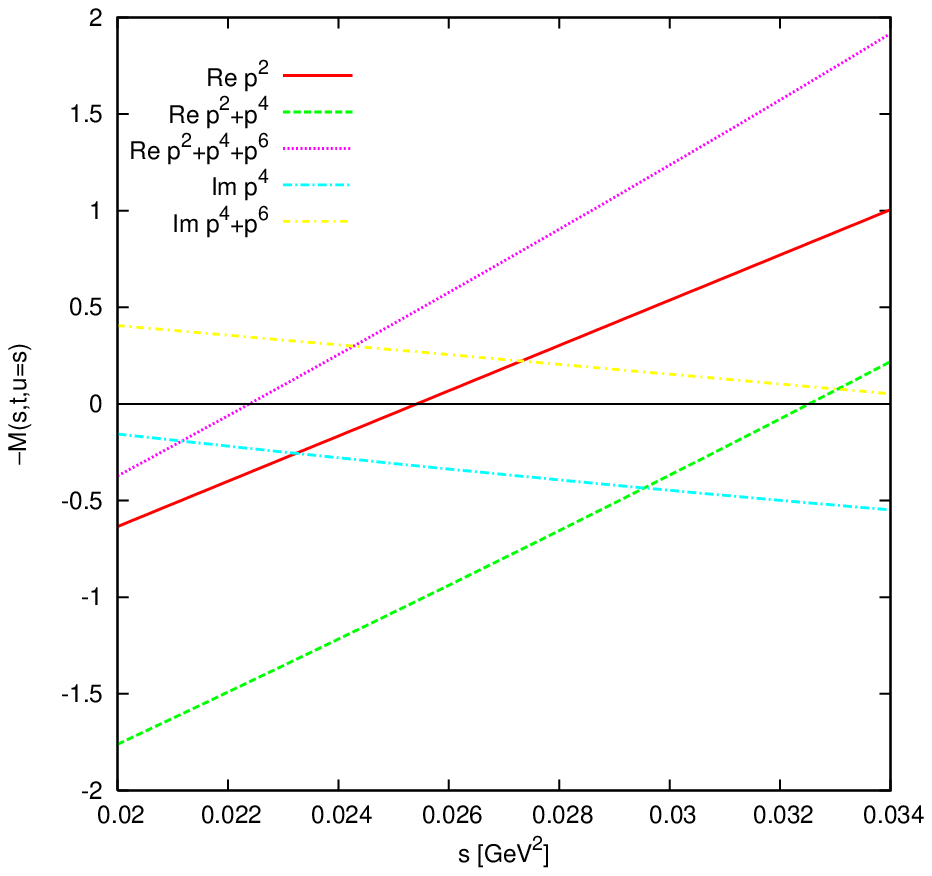}
\end{minipage}
\caption{$M(s,t,u)$ along the line with $s=u$ concentrated along
the Adler zero.}
\label{figadler}
} 
The current algebra prediction for the
Adler zero\footnote{We take here the zero of the real part of the amplitude as the Adler zero.}
is $s= \frac{4}{3} m_{\pi}^2$, independent of $t$.
The NLO effect produces a $t$ dependent shift in the zero for the real part.
A fairly large positive shift occurs along the line $t=u$ going from LO to NLO
and a smaller negative
shift from NLO to NNLO as can be seen in Fig.~\ref{figMstu2}(a).
Along the line $s=u$, a smaller positive shift appears from LO to NLO
at $s= 1.70~m_\pi^2$ shifting to $s= 1.17~m_\pi^2$ at NNLO.
The effects on the slopes can be judged from Figs.~\ref{figMstu2}(a)
and Fig.~\ref{figadler}.

\subsection{Dalitz Plot Parameters }
\label{sect:dalitz}

In general we use the data averages of the particle data group (PDG)\cite{PDG}.
For the distributions in the Dalitz plot the PDG presents no averages.
This distribution is usually described in terms of the variables
\ba
\label{defxy}
x &=& \sqrt3  \frac{T_{+} - T_{-}} {Q_{\eta}}
\nonumber\\
y &=& \frac{3T_{0}}{Q_{\eta}} -1  
\nonumber\\
Q_{\eta} &=& m_{\eta} - 2 m_{\pi^+} - m_{\pi^0}
\ea
for the charged decay. $T^i$ is the kinetic energy of pion $\pi^i$
in the final state. 
The standard parameterization of the
 Dalitz plot is (up to third order)
\be
\label{Dalitzparams}
|M|^2 = A_0^2\left( 1 + a y + b y^2  + d x^2 + f y^3 + g x^2 y+\cdots\right)\,.
\ee
Odd terms in $x$ are forbidden by charge conjugation,
all experimental results find them compatible with zero and
since all most precise experiments have presented fits with the odd terms
set to zero we use those. $f$ has only been measured by
KLOE \cite{KLOEcharged} 
and no experiment has attempted to determine $g$. Earlier experiments that 
only determined $a$ and $b$ are not included. 
The results are shown in Tab.~\ref{tabDalitzcharged}.
\TABLE{
\begin{tabular}{|c|ccc|}
\hline
Exp. & a & b & d  \\
\hline
\rule{0cm}{12pt}KLOE \cite{KLOEcharged} & $-1.090\pm0.005^{+0.008}_{-0.019}$ &
 $0.124\pm0.006\pm0.010$ & $0.057\pm0.006^{+0.007}_{-0.016}$\\
Crystal Barrel \cite{CBarrelcharged} & $-1.22\pm0.07$ & $0.22\pm0.11$ &
$0.06\pm0.04$ (input) \\
Layter et al. \cite{Layter73} & $-1.08\pm0.014$ & $0.034\pm0.027$ &
$0.046\pm0.031$ \\
Gormley et al. \cite{Gormley} &$-1.17\pm0.02$ & $0.21\pm0.03$ 
& $0.06\pm0.04$ \\
\hline
\end{tabular}
\caption{Measurements of the Dalitz plot distributions in 
$\eta\to\pi^+\pi^-\pi^0$. The parameters are defined in
Eq.~(\ref{Dalitzparams}).
The KLOE result \cite{KLOEcharged} for $f$ is $f=0.14\pm0.01\pm0.02$.
None of the others determined $f$.
 The Crystal Barrel used $d$ as input, but remarked that
$a$ and $b$ varied very little within the range of $d$ used. }
\label{tabDalitzcharged}
} 
There are discrepancies among data which are hard for us to discuss since
correlations are important. There is however a clear discrepancy between
KLOE\cite{KLOEcharged} and Gormley et al.\cite{Gormley} for $a$,
KLOE\cite{KLOEcharged}, Gormley et al.\cite{Gormley} and 
Layter et al.\cite{Layter73} have three rather incompatible numbers for $b$.
The results for $d$ are all compatible.

Similarly the neutral decay is parameterized by
\be
\label{Dalitzparamneutral}
|\overline M|^2 = \overline A_0^2\left( 1+2\alpha z+\cdots\right)\,.
\quad
z = \frac{2}{3}\sum_{i=1,3}
\left(\frac{3E_i-m_\eta}{m_\eta-3m_\pi^0}\right)^2\,.
\ee
$E_i$ is the energy of the $i$th pion in the final state.
The parameter $\alpha$ has been measured by several experiments. There are
two recent high precision measurements but they are in disagreement, however,
KLOE published a new analysis very recently.
\TABLE{
\begin{tabular}{|c|c|}
\hline
Exp. & $\alpha$\\
\hline
\rule{0cm}{12pt}
KLOE \cite{KLOEneutral2}      & $-0.027\pm0.004^{+0.004}_{-0.006}$\\
KLOE (prel)\cite{KLOEneutral1} & $-0.014\pm 0.005\pm0.004$ \\
Crystal Ball \cite{CBall} & $-0.031\pm0.004$ \\
WASA/CELSIUS \cite{WASAneutral} & $-0.026\pm0.010\pm0.010$ \\
Crystal Barrel \cite{CBarrelneutral} & $-0.052\pm0.017\pm0.010$ \\
GAMS2000 \cite{GAMS2000} & $-0.022\pm0.023$ \\
SND \cite{SND} & $-0.010 \pm 0.021 \pm 0.010$\\
\hline
\end{tabular}
\caption{Measurements of the Dalitz plot distribution in
$\eta\to\pi^0\pi^0\pi^0$. The parameter $\alpha$ is defined in
Eq.~(\ref{Dalitzparamneutral}).}
\label{tabDalitzneutral}
} 

Let us now discuss how to extract the Dalitz parameters 
in chiral perturbation theory. First for the charged decay.
The Dalitz plot variables $x$ and $y$ are related to the kinetic energy of the
outgoing particles. In terms of Mandelstam parameters they are given as,  
\ba
\label{defxy2}
x &=& 
\frac{\sqrt3}{2m_{\eta}Q_{\eta}} (u-t) 
\nonumber\\
y &=& 
\frac{3}{2m_{\eta}Q_{\eta}} \left( (m_{\eta} - m_{\pi^{o}} \right)^2 - s ) -1
\stackrel{\mathrm{iso}}{=} \frac{3}{2m_{\eta}Q_{\eta}}\left(s_0-s\right)\,.
\ea
The first equality is valid in general,
the second only in the isospin limit.
Our amplitudes are in the isospin limit, with a common pion mass
everywhere. The physical value gives
$Q_{\eta} = m_{\eta} - 2 m_{\pi^+} - m_{\pi^0} = 0.13318$.
For the pion mass we used, $3 m_\pi^2 = 2 m_{\pi^+}^2+m_{\pi^0}^2$, we
get $Q_\eta = 0.13313$~MeV. So the value of $Q_\eta$ is fine.
In the isospin limit, $x=y=0$ and $s=t=u=s_0$ coincide.
In the physical case there is a small difference,
$s=t=u=s_0$ corresponds to $y=-0.052$.

We  evaluate 
the decay amplitude $M(s,t,u)$ in the $s$-$t$ plane over the physical 
region. Since only two variables $s$ and $t$ are independent, 
the relation  $s+t+u = m_{\eta}^{2} + 2 m_{\pi^+}^{2} + m_{\pi^0}^{2}$  
is used to eliminate the third Mandelstam variable $u$ 
in the amplitude. We then fit (\ref{Dalitzparams}) to the amplitude
$|M|^2=|M^{(2)}+M^{(4)}+M^{(6)}|^2$ with as error $\Delta =
\mathrm{Re}M^{(6)}\mathrm{Re}(M^{(2)}+M^{(4)}+M^{(6)})+
\mathrm{Im}M^{(6)}\mathrm{Im}(M^{(2)}+M^{(4)}+M^{(6)})$. This is half of the NNLO
contribution.
The fitting error quoted in the result (\ref{dalitzNNLO})
is from this error, not
from errors in the input values of the $C_i^r$ and other LECs.
For $y$ we used the second expression in (\ref{defxy2}).
The fits have been performed to $|M^{(2)}|^2$, $|M^{(2)}+M^{(4)}|^2$
and  $|M^{(2)}+M^{(4)}+M^{(6)}|^2$ labeled LO, NLO and NNLO respectively.
In NNLO we have studied the effect of setting the $C_i^r=0$
and setting $L_i^r=C_i^r=0$ and in NLO for $L_i^r=0$.
In addition, we have checked how much changing $y$ from the second to the
first expression in (\ref{defxy2}) changes the results, labeled NNLOp
as well as including the $g$ term and the terms with $x^4,x^2y^2,y^4$,
labeled NNLOq.
\TABLE{
\begin{tabular}{|c|ccccc|}
\hline
               & $A_0^2$ & a & b & d & f \\
\hline
LO             & 120 & $-1.039$ & $0.270$ & $0.000$   & $0.000$ \\
NLO             & 314 & $-1.371$  & $0.452$ & $0.053$ & $0.027$\\
NLO ($L_i^r=0$) & 235 & $-1.263$& $0.407$ & $0.050$ & $0.015$\\
NNLO           & 538 &   $-1.271$ & $0.394$ & $0.055$ & $0.025$ \\
NNLOp          & 574 &   $-1.229$ & $0.366$ & $0.052$ & $0.023$ \\
NNLOq          & 535 & $-1.257$ & $0.397$ & $0.076$ & $0.004$ \\
NNLO ($\mu=0.6$~GeV) & 543 & $-1.300$ & $0.415$ & $0.055$ & $0.024$\\         
NNLO ($\mu=0.9$~GeV) & 548 & $-1.241$ & $0.374$ & $0.054$ & $0.025$\\         
NNLO ($C_i^r = 0$)& 465 & $-1.297$  & $0.404 $ & $0.058$ & $0.032$ \\
NNLO ($L_i^r=C_i^r = 0$)& 251 & $-1.241$  & $0.424$ & $0.050$ & $0.007$ \\
\hline
dispersive\cite{Kambor1} & --- & $-1.33$ & $0.26$ & $0.10$ & ---\\
tree dispersive\cite{BG} & --- & $-1.10$ & $0.33$ & $0.001$ & ---\\
absolute dispersive\cite{BG} & --- & $-1.21$ & $0.33$ & $0.04$ & ---\\
\hline
\end{tabular}
\caption{Theoretical estimate of the Dalitz plot distributions in 
$\eta\to\pi^+\pi^-\pi^0$. The parameters are defined in
Eq.~(\ref{Dalitzparams}). 
The line labeled NNLO is our central result.}
\label{tabDalitzcharged_theory}
} 
The results are given in Tab.~\ref{tabDalitzcharged_theory}.
The changes in going from NLO to NNLO are rather modest, the main change
is the overall normalization $A_0^2$. 
The linear slope $a$ is lowered somewhat but not enough to
reach the most recent experimental value. The same comment applies to the
quadratic slope $b$. $d$ is in agreement with the experimental values.
$f$ is always much smaller than the measured value of KLOE.
The results from the dispersive
calculations using Khuri-Treiman\cite{Kambor1}\footnote{Note that there
exists an analysis \cite{Martemyanov} using the method of \cite{Kambor1}
fitting the preliminary KLOE results. However, their final
normalization depends on the allowed ranges of parameters
given by \cite{Kambor1}. This is why we do not quote their numbers below.}
and the simplified
analysis of \cite{BG} with two different boundary conditions
are also shown in the table. \cite{Borasoy} used these parameters
as input and we have therefore not shown their result.
As final result we take the NNLO result of Tab.~\ref{tabDalitzcharged_theory}
with the MINUIT errors with inputs as above.
\ba
\label{dalitzNNLO}
A^2_0 &=& 538\pm 18\,,
\nonumber\\
a &=& -1.271\pm 0.075\,,
\nonumber\\
b &=& 0.394\pm 0.102\,,
\nonumber\\
d &=& 0.055\pm 0.057\,,
\nonumber\\
f &=& 0.025\pm0.160\,.
\ea

The fitting results for the neutral decay are shown in
Tab.~\ref{tabDalitzneutral_theory}. Again we have fitted LO, NLO, NNLO
and removed the contributions from $C_i^r$ and $L_i^r$ to see their effects.
The one labeled NNLOq is the same fit as NNLO but with the next two terms
in the expansion around $s=t=u=s_0$ included in the fit.
As in the case of the charged decay, one sees that going from NLO to NNLO
does not seem to change much in the Dalitz plot distributions but changes
mainly the overall normalization. It should be mentioned that because of the
three terms in (\ref{isorel}) large cancellations happen in the amplitude
for $\eta\to3\pi^0$. 
This, as well as an inequality between the slope parameters
is derived in App.~\ref{app:dalitz}. The inequality is always satisfied by our
results and the equality which results under additional assumptions
is reasonably well satisfied. 
\TABLE{
\begin{tabular}{|c|cc|}
\hline
    & $\overline A_0^2$ & $\alpha$\\
\hline
LO &  1090 & $ 0.000 $ \\
NLO & 2810 & $0.013$\\
NLO ($L_i^r=0$) & 2100 &   $ 0.016 $ \\
NNLO & 4790 & $ 0.013$ \\
NNLOq & 4790 & $ 0.014$ \\
NNLO ($C_i^r = 0$) & 4140 & $ 0.011$\\
NNLO ($L_i^r=C_i^r = 0$) & 2220 & $ 0.016$\\
\hline
dispersive\cite{Kambor1} & --- & $-(0.007$---$0.014)$\\
tree dispersive\cite{BG} & --- & $-0.0065$\\
absolute dispersive\cite{BG} & --- & $-0.007$\\
\cite{Borasoy}           & --- & $-0.031$\\
\hline
\end{tabular}
\caption{Theoretical estimates of the Dalitz plot distribution in
$\eta\to\pi^0\pi^0\pi^0$. The parameter $\alpha$ is defined in
Eq.~(\ref{Dalitzparamneutral}). The line labeled NNLO is the main result.}
\label{tabDalitzneutral_theory}
} 
Assuming the error on the NNLO result to be one half the NNLO contribution
as we did for the charged case leads to 
\ba
\overline A_0^2 &=& 4790\pm 160
\nonumber\\
\alpha &=& 0.013\pm0.032\,.
\ea
 The results from the dispersive
calculations using Khuri-Treiman\cite{Kambor1} and the simplified
analysis of \cite{BG} with two different boundary conditions
are also shown in the table. No theoretical prediction comes near
the experimental results of KLOE\cite{KLOEneutral2} and
Crystal Ball\cite{CBall} except for the result\footnote{That reference
combines a Chiral Lagrangian including the $\eta'$ and a model
for unitary resummations. Their good agreement with data is remarkable
but should also be understood using more controlled approximations.}
of \cite{Borasoy}.

It is clear from these results that further study on possible variations of
input parameters of ChPT is needed. One puzzling observation is however that
the main effect in the dispersive calculations is $S$-wave rescattering
and the input parameters used here give a very good prediction
for the scattering length $a_0^0$\cite{BDT1}.

\subsection{The ratio $r$
and decay rates}
\label{sect:decay}

We can now proceed to calculate the various decay rates from our amplitude.
Since the evaluation of the NNLO terms is very slow we have used
the following procedure. The actual shape of the
allowed $s,t,u$ region in the physical charged decay is somewhat different
from the one allowed with an equal pion mass. We therefore perform the
integration over the physically allowed values of
$s,t,u$ for the neutral and charged case.

The matrix-element $|M|^2$ is dealt with similar to the previous subsection.
First we fit $|M|^2$ with an expansion
in $x$ and $y$ up to fourth order in $x$ and $y$ for the charged and neutral
case. The values for $x$ and $y$ are calculated using the
values for $s,t,u$ with the formulas of the isospin limit
with the pion mass as defined above.

Doing that leads to the decay widths for the decays where we have indicated
the various orders
\ba
\label{resultGamma}
\Gamma(\eta\to\pi^+\pi^-\pi^0) &=& \sin^2\epsilon\,\cdot 
0.572~\mathrm{MeV\hskip1cm LO}\,,
\nonumber\\&&
\sin^2\epsilon\,\cdot 
1.59~\mathrm{MeV\hskip1cm NLO}\,,
\nonumber\\&&
\sin^2\epsilon\,\cdot 
2.68~\mathrm{MeV\hskip1cm NNLO}\,,
\nonumber\\&&
\sin^2\epsilon\,\cdot 
2.33~\mathrm{MeV\hskip1cm NNLO }\,C_i^r=0\,,
\nonumber\\
\Gamma(\eta\to\pi^0\pi^0\pi^0) &=& \sin^2\epsilon\,\cdot 
0.884~\mathrm{MeV\hskip1cm LO }\,,
\nonumber\\&&
\sin^2\epsilon\,\cdot 
2.31~\mathrm{MeV\hskip1cm NLO}\,,
\nonumber\\&&
\sin^2\epsilon\,\cdot 
3.94~\mathrm{MeV\hskip1cm NNLO}\,,
\nonumber\\&&
\sin^2\epsilon\,\cdot 
3.40~\mathrm{MeV\hskip1cm NNLO }\,C_i^r=0\,.
\ea

The numbers in (\ref{resultGamma})
can be used to calculate the ratio of decay rates
\be
r \equiv 
\frac{\Gamma(\eta\to\pi^0\pi^0\pi^0)}{\Gamma(\eta\to\pi^+\pi^-\pi^0)}\,.
\ee
To lowest order and neglecting the differences in phase space this
ratio is expected to be exactly 1.5. The correct treatment of
phasespace and the slightly different pion mass used lead to
\be
r_{\mathrm{LO}} = 1.54\,.
\ee
At higher orders a significant change is found with
\ba
r_{\mathrm{NLO}} &=& 1.46\,.
\nonumber\\
r_{\mathrm{NNLO}} &=& 1.47\,.
\nonumber\\
r_{\mathrm{NNLO }\,C_i^r=0} &=& 1.46\,.
\ea
The small changes from NLO to NNLO is because the higher order corrections
mainly change the overall size of the amplitude, not its shape.

This should be compared with the experimental result\cite{PDG}
of
\ba
r &=& 1.49\pm0.06\quad\mathrm{our~average}\,.
\nonumber\\
r &=& 1.43\pm0.04\quad\mathrm{our~fit}\,,
\ea
The different results are from the direct measurements or from the global
fit to eta decays. Our results agree excellently with each value.

\subsection{Discussion and the values of $R$ and $Q$}
\label{sect:RQ}

One of the prime reasons to study the hadronic eta decay to
pions is the direct determination of the double quark 
mass ratio, $Q^{2}$ defined as\cite{LeutwylerQM} 
\ba
Q^2 &=& \frac{m_{s}^{2} -\hat m^{2}}{m_{d}^{2} - m_{u}^2 } 
\ea
The reason for considering this quantity is that it is to first order
independent of a shift in the quark masses of the form
\ba
m_u&\to& m_u+\alpha m_d m_s\,,
\nonumber\\
m_d&\to& m_d+\alpha m_s m_u\,,
\nonumber\\
m_s&\to& m_s+\alpha m_u m_d\,.
\ea
Within the framework of standard ChPT such a change is unobservable,
it can always be compensated by a change in the values of
the LECs up to order $p^8$ effects. This was observed
to order $p^4$ in \cite{KM} but the generalization to higher orders
is obviously true. In our case, we are implicitly using assumptions
on the values of the constants such that this shift is fixed as discussed
in \cite{ABT4}. That is one of the reasons we pulled out an
overall factor of $\sin(\epsilon)$ rather than $Q^{-2}$ out of the
eta decay amplitude. To first order in isospin breaking we have
\be
\label{Repsilon}
\sin(\epsilon) = \epsilon = \frac{\sqrt{3}}{4}
\,\frac{m_d-m_u}{m_s-\hat m} =  \frac{\sqrt{3}}{4} \frac{1}{R}\,,
\ee
where the last equality defines the ratio $R$.

Since the process is $\eta\to3\pi$ is strongly protected from the
electromagnetic
interactions due to the chiral symmetry, we expect to obtain
the value $m_{d}-m_{u}$ rather well from this process. So let us see
what our results imply.
Using the experimental value\cite{PDG}
\be
\Gamma(\eta\to\pi^+\pi^-\pi^0) = 295\pm17~\mathrm{eV}
\ee
and (\ref{resultGamma}) and (\ref{Repsilon})
we obtain the values for $R$ quoted in Tab.~\ref{tabR})
in the line labeled $R(\eta)$.
\TABLE{
\begin{tabular}{|c|c|c|c|c|c|}
\hline
      & LO & NLO & NNLO & NNLO $(C_i^r=0)$\\
\hline
$R$ ($\eta$) &  19.1   & 31.8     & 42.2   &  38.7  \\
$R$ (Dashen) &  44     & 44 & 37 & ---\\
$R$ (Dashen-violation) & 36 & 37 & 32 & ---\\  
\hline
$Q$ ($\eta$) &  15.6   & 20.1  & 23.2   &  22.2  \\
$Q$ (Dashen) &  24     & 24 & 22 & ---\\
$Q$ (Dashen-violation) & 22 & 22 & 20 & ---\\  
\hline
\end{tabular}
\caption{The isospin breaking quantities R are evaluated
at $p^2$, $p^4$ and $p^6$. 
The $Q$ values are given with $Q^2\approx 12.7\,R$.
See text for details.}
\label{tabR}
} 
For comparison we also quote the values obtained at LO using
\be
R_{LO} = \frac{m_{K^0}^2+m_{K^+}^2-2 m_{\pi^0}^2}{2\left(m_{K^0}^2-m_{K^+}^2\right)}
\ee
where the QCD part of the masses should be included only.
The electromagnetic part of the $K^0$ and $\pi^0$ mass is taken to be
negligible and we use
\be
m_{K^+\mathrm{em}}^2 = x_D\left(m_{\pi^+}^2-m_{\pi^0}^2\right)\,.
\ee
We take $x_D=1$ in the line labeled Dashen\cite{Dashen}
and use the estimate $x_D=1.84$ \cite{BP} for the line labeled
Dashen-violation. The columns labeled NLO and NNLO quote the results
from \cite{ABT4} using the same input. These used $m_s/\hat m=24$
as input but changing it to 26 changes\footnote{This can be seen from
using the number for $m_u/m_d$ and $m_s/\hat m$ from Fig.~3(b)
in \cite{ABT4}.}
the value of $R$ by about half a unit.
Note that the limit $R<44$ \cite{LeutwylerQM} is satisfied by all the estimates
in Tab.~\ref{tabR}.

It is harder to draw conclusions on the value of $Q$. Its value
can be obtained from
\be
Q^2 =   \frac{m_{K}^{2}}
{m_{\pi}^{2} } R_{LO} (1 + \mathrm{NNLO}   ) \,,
\ee
where again only the QCD part of the masses should be included
and $m_K^2,m_\pi^2$ are the masses in the isospin limit.
The NNLO corrections are known \cite{ABT4} and not negligible.
However, they do depend at present on the input value of $m_s/\hat m$
used in the chiral fits, this can be seen in Fig.~3(a) of \cite{ABT4}.
Varying $m_s/\hat m$ from 24 to 26 changes Q by about one unit.
The analysis of \cite{LeutwylerQM} relied on the NNLO correction being small.

The relation between $Q^2$ and $R$ can be written as
\be
Q^2 = \frac{1}{2}\left(1+\frac{m_s}{\hat m}\right)\,R\,.
\ee
The analysis of \cite{LeutwylerQM} leads to a factor of about 12.7
using $m_s/\hat m=24.4$,
the input value for $m_s/\hat m=24$ to about 12.5 and for
 $m_s/\hat m=26$ to 13.5. After taking the square root, this range
corresponds to an uncertainty on $Q$ of about unit.
For completeness we have listed in Tab.~\ref{tabR} the values of $Q$ derived
from $R$ with the factor equal to 12.7.

\section{Conclusions}
\label{sect:conclusions
}
In this paper we have performed a NNLO calculation in standard ChPT
of the amplitude for $\eta\to3\pi$. This calculation was performed to
first order in isospin breaking and we have pulled out the overall factor
in the form of $\sin(\epsilon)$, the lowest order $\pi^0$-$\eta$ mixing angle.
The remainder of the amplitude is then dealt with in the isospin limit.
How we have dealt with the pion mass is described in Sect.~\ref{sect:firstlook}
and with the Dalitz plot distributions and decay rates in Sects.
\ref{sect:dalitz} and \ref{sect:decay}.

We find a reasonable enhancement over the NLO result of \cite{GL3}
which is somewhat larger than the estimates from the dispersive analyses
\cite{Kambor1,Anisovich1}. The shape of the amplitude is in better
agreement with \cite{Anisovich1} from comparing the published
plots along the line $s=u$. We have also commented on the position
of the Adler zero which was used in \cite{Anisovich1} to determine
their subtraction constants. The NNLO result for the Dalitz plot distributions
moves somewhat in the direction of the experimental results compared to
the NLO result but is not in good agreement. The same is also true
for the slope parameter $\alpha$ in the neutral decay. We always
obtain a positive value while experiment is consistently negative.
The amplitude for this decay has large cancellations and that
makes $\alpha$ a difficult parameter to predict, however all 
methods which include unitarity resummations and have published values
 \cite{BG,Kambor1,Borasoy} get a negative value. This we find puzzling,
since the main effect is $\pi\pi$ $S$-wave rescattering and our input values
give a good value for the scattering length $a_0^0$\cite{BDT1}.
We obtain very good agreement with the ratio $r$.

Since we find a somewhat larger enhancement of the decay rate then
\cite{Anisovich1} we also find a somewhat larger value for the isospin
breaking quantities $R$ and $Q$, which means that $m_d-m_u$ is somewhat
smaller than obtained in \cite{LeutwylerQM}. We also find values
that are not in agreement with the NNLO order fit to the meson
masses \cite{ABT4}.

The influence of changes in the input values is under study,
a first impression can be had from looking at the results for
the $C_i^r=0$ and the different choices of $\mu$.
A more detailed analysis is planned.

\acknowledgments

This work is supported in part by the European Commission RTN network,
Contract MRTN-CT-2006-035482  (FLAVIAnet), 
the European Community-Research Infrastructure
Activity Contract RII3-CT-2004-506078 (HadronPhysics) and
the Swedish Research Council.

\appendix

\section{A discussion on Dalitz plot parameters and the sign of $\alpha$}
\label{app:dalitz}

The goal of this appendix is to point a few simple observations
about relations between the Dalitzplot parameters.
We start by parameterinzing the amplitude for the charged decay
rate as
\ba
M(s,t,u) &=& A\left(1+\tilde a(s-s_0)+\tilde b(s-s_0)^2+\tilde d(u-t)^2+\cdots
\right)
\nonumber\\
&=& A\left(1+\overline a y +\overline b y^2+\overline d x^2+\cdots\right)
\ea
By computing $|M(s,t,u)|^2$
and using (\ref{defxy2}) this leads to the relations
\ba
\label{relslopes}
a &=& 2 \, \mathrm{Re}(\overline a)
=
-2 R_\eta\, \mathrm{Re}(\tilde a)\,,
\nonumber\\
b &=& |\overline a|^2+2 \mathrm{Re}(\overline b) =
 R_\eta^2\left(|\tilde a|^2+2 \mathrm{Re}(\tilde b)\right)\,,
\nonumber\\
d &=& 2\mathrm{Re}(\overline d) = 6 R_\eta^2 \,\mathrm{Re}(\tilde d)\,.
\ea
Here we defined $R_\eta = (2 m_\eta Q_\eta)/3$.
We can now use relation (\ref{isorel}) and $s+t+u=3s_0$. We obtain
\ba
\overline M(s,t,u) &\equiv& 
\overline A\left(1+\overline\alpha z+\cdots\right)
\nonumber\\
&=&
 A\left(3+\left(\tilde b+3\tilde d\right)
\left(\left(s-s_0\right)^2+\left(t-s_0\right)^2+\left(u-s_0\right)^2\right)
+\cdots\right)
\ea
Using the definition of $z$ these two can be related and give
\ba
\overline A = 3 A\,,
\ea
as well
\be
\alpha = \mathrm{Re}(\overline\alpha) = 
\frac{1}{2} R_\eta^2\,\mathrm{Re}\left(\tilde b+3\tilde d\right)
= \frac{1}{4}\left(d+b-R_\eta^2 |\tilde a|^2\right)
\,.
\ee
We thus obtain the relation
\be
\label{relalpha}
\alpha \le \frac{1}{4}\left(d+b-\frac{1}{4}a^2\right)\,.
\ee
Under the \emph{assumption} that $\mathrm{Im}(\tilde a)=0$, this turns
into an equality.
If one looks at the numbers in Tabs.~\ref{tabDalitzcharged}
and \ref{tabDalitzcharged_theory} we see that there is a very strong
cancellation on the r.h.s. of (\ref{relalpha}).
Note that the KLOE results satisfy
the relation with equality
quite well and the theory results satisfy it within 30\% or so. The inequality
is satisfied in all cases.
The underlying reason why it is difficult to get a negative $\alpha$ seems to
be that the value obtained for $b$ is too large compared to experiment.

For our NNLO result we can perform the fit also to the amplitude directly
and we obtain
\ba
A &=& -22.7-i\,4.38\,,
\nonumber\\
\overline a &=& -0.631-i 0.183\,,
\nonumber\\
\overline b &=& -0.017+i\,0.025\,,
\nonumber\\
\overline d &=& 0.040-i0.023\,.
\ea
We see that the relation (\ref{relslopes}) is satisfied.
In general, $\overline a$ is sizable but $\overline b$, $\overline c$, and the
higher order in $x,y$ generalizations are small.

\section{The order $p^4$ expression}
\label{App:p4}

In this appendix we quote the precise way in which we have defined the
order $p^4$ contribution. It agrees with the result of \cite{GL3}.
We use the notation
\be
P_{\eta\pi} = m_\eta^2-m_{\pi^0}^2\,.
\ee
This appears after $\Delta m_1^2$ and $\Delta m_2^2$ of (\ref{fullp4})
have been rewritten in the physical masses.

The functions are defined in (\ref{defMi4}).
\ba
\lefteqn{
M^{(4)}_0(s) =
       + L^r_{8}\, \Big( 32\,m_{K^0}^2\,s - 32\,m_{\pi^0}^2\,s - 256/3\,m_{\pi^0}^2\,m_{K^0}^2 + 256/3\,m_{\pi^0}^4 \Big)
}
\nonumber\\&&
       + L^r_{7}\, \Big( 96\,m_{K^0}^2\,s - 96\,m_{\pi^0}^2\,s - 640/3\,m_{\pi^0}^2\,m_{K^0}^2 + 640/3\,m_{\pi^0}^4 \Big)
\nonumber\\&&
       + L^r_{5}\, \Big(  - 8\,m_{\pi^0}^2\,s + 64/9\,m_{\pi^0}^2\,m_{K^0}^2 + 32/9\,m_{\pi^0}^4 \Big)
\nonumber\\&&
       + L^r_{3}\, \Big(  - 16/9\,s^2 + 16/3\,m_{K^0}^2\,s - 64/27\,m_{K^0}^4
 + 32/3\,m_{\pi^0}^2\,s
 - 256/27\,m_{\pi^0}^2\,m_{K^0}^2 - 256/27\,m_{\pi^0}^4 \Big)
\nonumber\\&&
       + \frac{1}{16\pi^2} \, \Big( 1/9\,s^2 + 1/6\,m_{K^0}^2\,s - 2/27\,m_{K^0}^4 + 1/3\,m_{\pi^0}^2\,s - 8/27\,m_{\pi^0}^2\,
         m_{K^0}^2 - 8/27\,m_{\pi^0}^4 \Big)
\nonumber\\&&
       + \overline{A}(m_{\pi^0}^2)\,P_{\eta \pi}^{-1} \, \Big( 4/3\,m_{\pi^0}^2\,s - 16/9\,m_{\pi^0}^4 \Big)
       + \overline{A}(m_{\pi^0}^2) \, \Big(  - 3/2\,s + 4/9\,m_{K^0}^2 + 20/9\,m_{\pi^0}^2 \Big)
\nonumber\\&&
       + \overline{A}(m_{K^0}^2)\,P_{\eta \pi}^{-1} \, \Big(  - 4/3\,m_{\pi^0}^2\,s + 16/9\,m_{\pi^0}^4 \Big)
       + \overline{A}(m_{K^0}^2) \, \Big(  - 1/2\,s + 2/9\,m_{K^0}^2 - 2/9\,m_{\pi^0}^2 \Big)
\nonumber\\&&
       + \overline{A}(m_\eta^2) \, \Big( 1/2\,s - 2/3\,m_{\pi^0}^2 \Big)
\nonumber\\&&
       + \overline{B}(m_{\pi^0}^2,m_{\pi^0}^2,s) \, \Big(  - 2/3\,s^2 - 4/9\,m_{K^0}^2\,s + 5/3\,m_{\pi^0}^2\,s + 2/9\,
         m_{\pi^0}^2\,m_{K^0}^2 - 2/3\,m_{\pi^0}^4 \Big)
\nonumber\\&&
       + \overline{B}(m_{\pi^0}^2,m_\eta^2,s) \, \Big(  - 2/9\,m_{\pi^0}^2\,s - 4/27\,m_{\pi^0}^2\,m_{K^0}^2 + 4/9\,m_{\pi^0}^4 \Big)
\nonumber\\&&
       + \overline{B}(m_{K^0}^2,m_{K^0}^2,s) \, \Big( 1/2\,s^2 - 2/3\,m_{K^0}^2\,s + 4/9\,m_{K^0}^4 - 1/3\,m_{\pi^0}^2
         \,s \Big)
\nonumber\\&&
       + \overline{B}(m_{K^0}^2,m_{K^0}^2,0) \, \Big( 3/2\,m_{K^0}^2\,s - 2/3\,m_{K^0}^4 - 1/2\,m_{\pi^0}^2\,s - 2/3\,
\nonumber\\&&
         m_{\pi^0}^2\,m_{K^0}^2 \Big)
\nonumber\\&&
       + \overline{B}(m_\eta^2,m_\eta^2,s) \, \Big( 2/9\,m_{\pi^0}^2\,m_{K^0}^2 - 2/9\,m_{\pi^0}^4 \Big)
\nonumber\\&&
       + \overline{C}(m_{K^0}^2,m_{K^0}^2,m_{K^0}^2,s) \, \Big(  - 1/2\,m_{K^0}^2\,s^2 + 2/3\,m_{K^0}^4\,s + 1/2\,
         m_{\pi^0}^2\,s^2 - 2/3\,m_{\pi^0}^2\,m_{K^0}^2\,s \Big)
\nonumber\\&&
\ea

\be
M^{(4)}_1(t) =
      \overline{B}(m_{\pi^0}^2,m_{\pi^0}^2,t) \, \Big(  - 1/12\,t + 1/3\,m_{\pi^0}^2 \Big)
       + \overline{B}(m_{K^0}^2,m_{K^0}^2,t) \, \Big(  - 1/24\,t + 1/6\,m_{K^0}^2 \Big)\,.
\ee

\ba
\lefteqn{M^{(4)}_2(t) =
       L^r_{3}\, \Big( 4/3\,t^2 \Big)
       + \frac{1}{16\pi^2} \, \Big(  - 1/12\,t^2 \Big)
}
\nonumber\\&&
       + \overline{B}(m_{\pi^0}^2,m_{\pi^0}^2,t) \, \Big(  - 1/4\,t^2 + 1/3\,m_{K^0}^2\,t + 1/2\,m_{\pi^0}^2\,t - 2/3\,
         m_{\pi^0}^2\,m_{K^0}^2 \Big)
\nonumber\\&&
       + \overline{B}(m_{\pi^0}^2,m_\eta^2,t) \, \Big( 1/6\,m_{\pi^0}^2\,t - 2/9\,m_{\pi^0}^2\,m_{K^0}^2 \Big)
       + \overline{B}(m_{K^0}^2,m_{K^0}^2,t) \, \Big( 3/8\,t^2 - m_{K^0}^2\,t + 2/3\,m_{K^0}^4 \Big)\,.
\nonumber\\&&
\ea

\section{The order $p^6$ LECs dependent part}
\label{App:p6}

In this appendix we give the amplitude dependence on the order $p^6$ LECs
 $C_i^r$ in the form of $M_I^C(t)$ as defined (\ref{defMi6parts}).
\ba
\lefteqn{ M_0^C(s) =
       + C^r_{33} \, \Big( 256\,m_{K^0}^4\,s - 512/9\,m_{K^0}^6 - 128\,m_{\pi^0}^2\,m_{K^0}^2\,s - 1280/9\,m_{\pi^0}^2\,
         m_{K^0}^4 - 128\,m_{\pi^0}^4\,s
}
\nonumber\\&&
 - 5120/9\,m_{\pi^0}^4\,m_{K^0}^2 + 768\,m_{\pi^0}^6 \Big)
\nonumber\\&&
       + C^r_{32} \, \Big( 128\,m_{K^0}^4\,s - 512/9\,m_{K^0}^6 - 64\,m_{\pi^0}^2\,m_{K^0}^2\,s - 1280/9\,m_{\pi^0}^2\,
         m_{K^0}^4
\nonumber\\&&
 - 64\,m_{\pi^0}^4\,s - 1280/9\,m_{\pi^0}^4\,m_{K^0}^2 + 1024/3\,m_{\pi^0}^6 \Big)
\nonumber\\&&
       + C^r_{31} \, \Big( 128\,m_{K^0}^4\,s - 512/9\,m_{K^0}^6 - 64\,m_{\pi^0}^2\,m_{K^0}^2\,s - 256/3\,m_{\pi^0}^2\,
         m_{K^0}^4
\nonumber\\&&
 - 64\,m_{\pi^0}^4\,s - 1024/3\,m_{\pi^0}^4\,m_{K^0}^2 + 4352/9\,m_{\pi^0}^6 \Big)
\nonumber\\&&
       + C^r_{29} \, \Big( 128/3\,m_{\pi^0}^2\,m_{K^0}^2\,s + 64/3\,m_{\pi^0}^4\,s - 256/9\,m_{\pi^0}^4\,m_{K^0}^2 - 512/
         9\,m_{\pi^0}^6 \Big)
\nonumber\\&&
       + C^r_{28} \, \Big( 256/3\,m_{K^0}^4\,s - 512/3\,m_{\pi^0}^2\,m_{K^0}^2\,s - 512/3\,m_{\pi^0}^2\,m_{K^0}^4 + 256/
         3\,m_{\pi^0}^4\,s
\nonumber\\&&
 + 1024/3\,m_{\pi^0}^4\,m_{K^0}^2 - 512/3\,m_{\pi^0}^6 \Big)
\nonumber\\&&
       + C^r_{27} \, \Big( 128/3\,m_{K^0}^4\,s - 256/3\,m_{\pi^0}^2\,m_{K^0}^2\,s - 512/9\,m_{\pi^0}^2\,m_{K^0}^4 + 128/
         3\,m_{\pi^0}^4\,s
\nonumber\\&&
 + 1792/9\,m_{\pi^0}^4\,m_{K^0}^2 - 1280/9\,m_{\pi^0}^6 \Big)
\nonumber\\&&
       + C^r_{26} \, \Big(  - 64/3\,m_{\pi^0}^2\,m_{K^0}^2\,s + 256/9\,m_{\pi^0}^2\,m_{K^0}^4 - 32/3\,m_{\pi^0}^4\,s +
         128/3\,m_{\pi^0}^4\,m_{K^0}^2 - 256/9\,m_{\pi^0}^6 \Big)
\nonumber\\&&
       + C^r_{25} \, \Big(  - 32/3\,m_{K^0}^2\,s^2 + 64/9\,m_{K^0}^4\,s - 32/9\,m_{\pi^0}^2\,s^2 + 256/9\,
         m_{\pi^0}^2\,m_{K^0}^2\,s - 256/27\,m_{\pi^0}^2\,m_{K^0}^4
\nonumber\\&&
 + 256/9\,m_{\pi^0}^4\,s - 1024/27\,m_{\pi^0}^4\,
         m_{K^0}^2 - 1024/27\,m_{\pi^0}^6 \Big)
\nonumber\\&&
       + C^r_{24} \, \Big(  - 160/9\,m_{K^0}^2\,s^2 - 64/3\,m_{K^0}^4\,s + 512/27\,m_{K^0}^6 + 160/9\,
         m_{\pi^0}^2\,s^2
\nonumber\\&&
 + 128/3\,m_{\pi^0}^2\,m_{K^0}^2\,s + 256/9\,m_{\pi^0}^2\,m_{K^0}^4 - 64/3\,m_{\pi^0}^4\,s -
         256/9\,m_{\pi^0}^4\,m_{K^0}^2 - 512/27\,m_{\pi^0}^6 \Big)
\nonumber\\&&
       + C^r_{22} \, \Big(  - 64/9\,m_{K^0}^2\,s^2 - 256/9\,m_{K^0}^4\,s + 512/27\,m_{K^0}^6 + 64/3\,
         m_{\pi^0}^2\,s^2
\nonumber\\&&
 + 128/9\,m_{\pi^0}^2\,m_{K^0}^2\,s + 1024/27\,m_{\pi^0}^2\,m_{K^0}^4 - 448/9\,m_{\pi^0}^4\,s
          + 256/27\,m_{\pi^0}^4\,m_{K^0}^2 + 512/27\,m_{\pi^0}^6 \Big)
\nonumber\\&&
       + C^r_{21} \, \Big(  - 256/3\,m_{K^0}^6 + 64\,m_{\pi^0}^4\,m_{K^0}^2 + 64/3\,m_{\pi^0}^6 \Big)
\nonumber\\&&
       + C^r_{20} \, \Big( 128\,m_{K^0}^4\,s - 256/3\,m_{K^0}^6 - 64\,m_{\pi^0}^2\,m_{K^0}^2\,s - 256/3\,m_{\pi^0}^2\,
         m_{K^0}^4 - 64\,m_{\pi^0}^4\,s
\nonumber\\&&
 - 192\,m_{\pi^0}^4\,m_{K^0}^2 + 1088/3\,m_{\pi^0}^6 \Big)
\nonumber\\&&
       + C^r_{19} \, \Big( 192\,m_{K^0}^4\,s - 256/3\,m_{K^0}^6 - 96\,m_{\pi^0}^2\,m_{K^0}^2\,s - 128\,m_{\pi^0}^2\,
         m_{K^0}^4 - 96\,m_{\pi^0}^4\,s
\nonumber\\&&
 - 320\,m_{\pi^0}^4\,m_{K^0}^2 + 1600/3\,m_{\pi^0}^6 \Big)
\nonumber\\&&
       + C^r_{18} \, \Big(  - 320/3\,m_{K^0}^4\,s + 512/27\,m_{K^0}^6 + 352/3\,m_{\pi^0}^2\,m_{K^0}^2\,s + 256/
         3\,m_{\pi^0}^2\,m_{K^0}^4 - 32/3\,m_{\pi^0}^4\,s
\nonumber\\&&
 - 128/9\,m_{\pi^0}^4\,m_{K^0}^2 - 2432/27\,m_{\pi^0}^6 \Big)
\nonumber\\&&
       + C^r_{17} \, \Big(  - 64/3\,m_{K^0}^4\,s + 512/27\,m_{K^0}^6 + 32\,m_{\pi^0}^2\,m_{K^0}^2\,s - 256/27\,
         m_{\pi^0}^2\,m_{K^0}^4 - 80/3\,m_{\pi^0}^4\,s
\nonumber\\&&
 + 1472/27\,m_{\pi^0}^4\,m_{K^0}^2 - 128/3\,m_{\pi^0}^6 \Big)
\nonumber\\&&
       + C^r_{16} \, \Big( 448/3\,m_{K^0}^4\,s + 256/9\,m_{K^0}^6 - 704/3\,m_{\pi^0}^2\,m_{K^0}^2\,s - 2816/9\,
         m_{\pi^0}^2\,m_{K^0}^4 + 400/3\,m_{\pi^0}^4\,s
\nonumber\\&&
 + 4288/9\,m_{\pi^0}^4\,m_{K^0}^2 - 256\,m_{\pi^0}^6 \Big)
\nonumber\\&&
       + C^r_{15} \, \Big( 512/27\,m_{K^0}^6 + 256/9\,m_{\pi^0}^2\,m_{K^0}^4 - 256/9\,m_{\pi^0}^4\,m_{K^0}^2 - 512/
         27\,m_{\pi^0}^6 \Big)
\nonumber\\&&
       + C^r_{14} \, \Big(  - 64/3\,m_{K^0}^4\,s + 512/27\,m_{K^0}^6 - 32\,m_{\pi^0}^2\,m_{K^0}^2\,s + 512/27\,
         m_{\pi^0}^2\,m_{K^0}^4 + 112/3\,m_{\pi^0}^4\,s
\nonumber\\&&
 + 2240/27\,m_{\pi^0}^4\,m_{K^0}^2 - 896/9\,m_{\pi^0}^6 \Big)
\nonumber\\&&
       + C^r_{13} \, \Big(  - 128/3\,m_{K^0}^2\,s^2 + 128\,m_{K^0}^4\,s - 512/27\,m_{K^0}^6 - 64/3\,m_{\pi^0}^2
         \,s^2 + 320\,m_{\pi^0}^2\,m_{K^0}^2\,s
\nonumber\\&&
 - 1792/9\,m_{\pi^0}^2\,m_{K^0}^4 + 128\,m_{\pi^0}^4\,s - 3584/9\,
         m_{\pi^0}^4\,m_{K^0}^2 - 4096/27\,m_{\pi^0}^6 \Big)
\nonumber\\&&
       + C^r_{12} \, \Big( 64/9\,m_{K^0}^2\,s^2 + 512/9\,m_{K^0}^4\,s + 512/81\,m_{K^0}^6 - 64/3\,m_{\pi^0}^2\,
         s^2 - 256/9\,m_{\pi^0}^2\,m_{K^0}^2\,s
\nonumber\\&&
 - 1024/9\,m_{\pi^0}^2\,m_{K^0}^4 + 896/9\,m_{\pi^0}^4\,s + 1792/
         27\,m_{\pi^0}^4\,m_{K^0}^2 - 10496/81\,m_{\pi^0}^6 \Big)
\nonumber\\&&
       + C^r_{11} \, \Big( 128/9\,m_{K^0}^2\,s^2 - 128/3\,m_{K^0}^4\,s + 512/27\,m_{K^0}^6 + 64/9\,m_{\pi^0}^2\,
         s^2 - 320/3\,m_{\pi^0}^2\,m_{K^0}^2\,s
\nonumber\\&&
 + 256/3\,m_{\pi^0}^2\,m_{K^0}^4 - 128/3\,m_{\pi^0}^4\,s + 1024/9
         \,m_{\pi^0}^4\,m_{K^0}^2 + 1024/27\,m_{\pi^0}^6 \Big)
\nonumber\\&&
       + C^r_{10} \, \Big(  - 32/9\,m_{K^0}^2\,s^2 - 64/9\,m_{K^0}^4\,s + 256/27\,m_{K^0}^6 + 32/3\,m_{\pi^0}^2
         \,s^2 - 256/9\,m_{\pi^0}^2\,m_{K^0}^2\,s
\nonumber\\&&
 + 256/27\,m_{\pi^0}^2\,m_{K^0}^4 - 256/9\,m_{\pi^0}^4\,s + 1408/
         27\,m_{\pi^0}^4\,m_{K^0}^2 + 128/9\,m_{\pi^0}^6 \Big)
\nonumber\\&&
       + C^r_{9} \, \Big(  - 160/9\,m_{K^0}^2\,s^2 + 512/27\,m_{K^0}^6 + 160/9\,m_{\pi^0}^2\,s^2 - 256/9\,
         m_{\pi^0}^2\,m_{K^0}^4 + 128/3\,m_{\pi^0}^4\,m_{K^0}^2
\nonumber\\&&
 - 896/27\,m_{\pi^0}^6 \Big)
\nonumber\\&&
       + C^r_{8} \, \Big(  - 32/9\,m_{K^0}^2\,s^2 - 64/9\,m_{K^0}^4\,s + 256/27\,m_{K^0}^6 + 32/9\,m_{\pi^0}^2\,
         m_{K^0}^2\,s - 128/27\,m_{\pi^0}^2\,m_{K^0}^4 
\nonumber\\&&
+ 320/9\,m_{\pi^0}^4\,s - 128/27\,m_{\pi^0}^4\,m_{K^0}^2 -
         128/3\,m_{\pi^0}^6 \Big)
\nonumber\\&&
       + C^r_{6} \, \Big(  - 64/9\,m_{K^0}^2\,s^2 + 64/3\,m_{K^0}^4\,s - 256/27\,m_{K^0}^6 - 32/9\,m_{\pi^0}^2\,
         s^2 + 160/3\,m_{\pi^0}^2\,m_{K^0}^2\,s
\nonumber\\&&
 - 128/3\,m_{\pi^0}^2\,m_{K^0}^4 + 64/3\,m_{\pi^0}^4\,s - 512/9\,
         m_{\pi^0}^4\,m_{K^0}^2 - 512/27\,m_{\pi^0}^6 \Big)
\nonumber\\&&
       + C^r_{5} \, \Big(  - 32/3\,m_{K^0}^2\,s^2 + 128/9\,m_{K^0}^4\,s + 64/9\,m_{\pi^0}^2\,s^2 + 224/9\,
         m_{\pi^0}^2\,m_{K^0}^2\,s - 896/27\,m_{\pi^0}^2\,m_{K^0}^4
\nonumber\\&&
 - 64/9\,m_{\pi^0}^4\,s - 128/27\,m_{\pi^0}^4\,m_{K^0}^2
          - 128/27\,m_{\pi^0}^6 \Big)
\nonumber\\&&
       + C^r_{4} \, \Big(  - 64/9\,s^3 + 128/27\,m_{K^0}^2\,s^2 + 128/9\,m_{K^0}^4\,s - 512/81\,
         m_{K^0}^6 + 256/27\,m_{\pi^0}^2\,s^2
\nonumber\\&&
 - 256/9\,m_{\pi^0}^2\,m_{K^0}^2\,s + 128/9\,m_{\pi^0}^4\,s + 512/
         27\,m_{\pi^0}^4\,m_{K^0}^2 - 1024/81\,m_{\pi^0}^6 \Big)
\nonumber\\&&
       + C^r_{3} \, \Big(  - 64/9\,s^3 + 128/9\,m_{K^0}^2\,s^2 - 128/9\,m_{K^0}^4\,s + 512/81\,m_{K^0}^6
          + 256/9\,m_{\pi^0}^2\,s^2
\nonumber\\&&
 - 896/9\,m_{\pi^0}^2\,m_{K^0}^2\,s
\nonumber\\&&
 + 512/9\,m_{\pi^0}^2\,m_{K^0}^4 - 704/9\,
         m_{\pi^0}^4\,s + 3328/27\,m_{\pi^0}^4\,m_{K^0}^2 + 5632/81\,m_{\pi^0}^6 \Big)
\nonumber\\&&
       + C^r_{1} \, \Big( 32/9\,s^3 - 64/9\,m_{K^0}^2\,s^2 + 64/9\,m_{K^0}^4\,s - 256/81\,m_{K^0}^6 -
         128/9\,m_{\pi^0}^2\,s^2
\nonumber\\&&
 + 448/9\,m_{\pi^0}^2\,m_{K^0}^2\,s - 256/9\,m_{\pi^0}^2\,m_{K^0}^4 + 352/9\,m_{\pi^0}^4
         \,s - 1664/27\,m_{\pi^0}^4\,m_{K^0}^2 - 2816/81\,m_{\pi^0}^6 \Big)
\nonumber\\&&
\ea

\be
M^C_1(t) = 0\,.
\ee
\ba
\lefteqn{ M^C_2(t) =
        C^r_{25} \, \Big( 32/3\,m_{\pi^0}^2\,t^2 \Big)
       + C^r_{24} \, 
          \Big(  - 32/3\,m_{K^0}^2\,t^2 + 32/3\,m_{\pi^0}^2\,t^2 \Big)
       + C^r_{22} \, \Big(  - 32/3\,m_{K^0}^2\,t^2 \Big)
}
\nonumber\\&&
       + C^r_{13} \, \Big( 32\,m_{K^0}^2\,t^2 + 16\,m_{\pi^0}^2\,t^2 \Big)
       + C^r_{12} \, \Big( 32/3\,m_{K^0}^2\,t^2 \Big)
\nonumber\\&&
       + C^r_{11} \, \Big(  - 32/3\,m_{K^0}^2\,t^2 - 16/3\,m_{\pi^0}^2\,t^2 \Big)
       + C^r_{10} \, \Big(  - 16/3\,m_{K^0}^2\,t^2 \Big)
\nonumber\\&&
       + C^r_{9} \, \Big(  - 32/3\,m_{K^0}^2\,t^2 + 32/3\,m_{\pi^0}^2\,t^2 \Big)
       + C^r_{8} \, \Big(  - 16/3\,m_{K^0}^2\,t^2 + 8\,m_{\pi^0}^2\,t^2 \Big)
\nonumber\\&&
       + C^r_{6} \, \Big( 16/3\,m_{K^0}^2\,t^2 + 8/3\,m_{\pi^0}^2\,t^2 \Big)
       + C^r_{5} \, \Big( 8/3\,m_{\pi^0}^2\,t^2 \Big)
\nonumber\\&&
       + C^r_{4} \, \Big( 16/3\,t^3 - 32/9\,m_{K^0}^2\,t^2 - 64/9\,m_{\pi^0}^2\,t^2 \Big)
       + C^r_{3} \, \Big( 16/3\,t^3 - 32/3\,m_{K^0}^2\,t^2 - 64/3\,m_{\pi^0}^2\,t^2 \Big)
\nonumber\\&&
       + C^r_{1} \, \Big(  - 8/3\,t^3 + 16/3\,m_{K^0}^2\,t^2 + 32/3\,m_{\pi^0}^2\,t^2 \Big)
\ea

\end{document}